\documentclass{emulateapj}
\usepackage{lscape,color}
\slugcomment{resubmitted to AJ on 8 October 2009}
\shorttitle{Far-IR spectra of compact sources in the LMC}
\shortauthors{van Loon et al.}
\begin{document}
\title{A {\it Spitzer Space Telescope} far-infrared spectral atlas of compact
sources in the Magellanic Clouds. I. The Large Magellanic Cloud}
\author{Jacco Th. van Loon\altaffilmark{1}}
\email{jacco@astro.keele.ac.uk}
\author{Joana M. Oliveira\altaffilmark{1}}
\author{Karl D. Gordon\altaffilmark{2}}
\author{Margaret Meixner\altaffilmark{2}}
\author{Bernie Shiao\altaffilmark{2}}
\author{Martha L. Boyer\altaffilmark{2}}
\author{F. Kemper\altaffilmark{3}}
\author{Paul M. Woods\altaffilmark{3}}
\author{A. G. G. M. Tielens\altaffilmark{4}}
\author{Massimo Marengo\altaffilmark{5,6}}
\author{Remy Indebetouw\altaffilmark{7,8}}
\author{G. C. Sloan\altaffilmark{9}}
\author{C.-H. Rosie Chen\altaffilmark{7}}
\affil{
$^{1}$ Astrophysics Group, Lennard-Jones Laboratories, Keele University,
Staffordshire ST5 5BG, UK\\
$^{2}$ Space Telescope Science Institute, 3700 San Martin Drive, Baltimore, MD
21218, USA\\
$^{3}$ Jodrell Bank Centre for Astrophysics, Alan Turing Building, School of
Physics and Astronomy, The University of Manchester, Oxford Road, Manchester
M13 9PL, UK\\
$^{4}$ Leiden Observatory, P.O.\ Box 9513, NL-2300 RA Leiden, The
Netherlands\\
$^{5}$ Harvard-Smithsonian Center for Astrophysics, 60 Garden Street,
Cambridge, MA 02138, USA\\
$^{6}$ Department of Physics and Astronomy, Iowa State University, Ames IA,
USA\\
$^{7}$ Department of Astronomy, University of Virginia, P.O.\ Box 400325,
Charlottesville, VA 22904, USA\\
$^{8}$ National Radio Astronomy Observatory, 520 Edgemont Road,
Charlottesville, VA 22903, USA\\
$^{9}$ Department of Astronomy, Cornell University, Ithaca, NY 14853, USA
}
\begin{abstract}
We present far-infrared spectra, $\lambda$=52--93 $\mu$m, obtained with the
{\it Spitzer Space Telescope} in the Spectral Energy Distribution mode of its
MIPS instrument, of a representative sample of the most luminous compact
far-infrared sources in the Large Magellanic Cloud. These include carbon
stars, OH/IR Asymptotic Giant Branch (AGB) stars, post-AGB objects and
Planetary Nebulae, the R\,CrB-type star HV\,2671, the OH/IR red supergiants
WOH\,G064 and IRAS\,05280$-$6910, the three B[e] stars IRAS\,04530$-$6916,
R\,66 and R\,126, the Wolf-Rayet star Brey\,3a, the Luminous Blue Variable
(LBV) R\,71, the supernova remnant N\,49, a large number of young stellar
objects (YSOs), compact H\,{\sc ii} regions and molecular cores, and a
background galaxy at a redshift $z\simeq0.175$. We use the spectra to
constrain the presence and temperature of cold dust and the excitation
conditions and shocks within the neutral and ionized gas, in the circumstellar
environments and interfaces with the surrounding interstellar medium (ISM).
First, we introduce a spectral classification scheme. Then, we measure line
strengths, dust temperatures, and IR luminosities. Objects associated with
star formation are readily distinguished from evolved stars by their cold dust
and/or fine-structure lines. Evolved stars, including the LBV R\,71, lack cold
dust except in some cases where we argue that this is swept-up ISM. This leads
to an estimate of the duration of the prolific dust-producing phase
(``superwind'') of several thousand years for both RSGs and massive AGB stars,
with a similar fractional mass loss experienced despite the different masses.
We tentatively detect line emission from neutral oxygen in the extreme RSG
WOH\,G064, which suggests a large dust-free cavity with implications for the
wind driving. In N\,49, the shock between the supernova ejecta and ISM is
revealed in spectacular fashion by its strong [O\,{\sc i}] $\lambda$63-$\mu$m
emission and possibly water vapour; we estimate that 0.2 M$_\odot$ of ISM dust
was swept up. On the other hand, some of the compact H\,{\sc ii} regions
display pronounced [O\,{\sc iii}] $\lambda$88-$\mu$m emission. The efficiency
of photo-electric heating in the interfaces of ionized gas and molecular
clouds is estimated at 0.1--0.3\%. We confirm earlier indications of a low
nitrogen content in the LMC. Evidence for solid state emission features is
found in both young and evolved objects, but the carriers of these features
remain elusive; some of the YSOs are found to contain crystalline water ice.
The spectra constitute a valuable resource for the planning and interpretation
of observations with the {\it Herschel Space Observatory} and the {\it
Stratospheric Observatory For Infrared Astronomy} (SOFIA).
\end{abstract}

\section{Introduction}

About the cycle of gas and dust that drives galaxy evolution, much can be
learnt from the interfaces between the sources of feedback and the
interstellar medium (ISM), and between the ISM and the dense cores of
molecular clouds wherein new generations of stars may form. These regions are
characterized by the cooling ejecta from evolved stars and supernovae, and
clouds heated by the radiation and shocks from hot stars, in supernova
remnants (SNRs) and young stellar objects (YSOs) embedded in molecular clouds.

The interaction regions with the ISM lend themselves particularly well to
investigation in the infrared (IR) domain, notably in the 50--100 $\mu$m
region; cool dust ($\sim20$--100 K) shines brightly at these wavelengths, and
several strong atomic and ionic transitions of abundant elements (viz.\
[O\,{\sc i}] at $\lambda=63$ $\mu$m, [O\,{\sc iii}] at $\lambda=88$ $\mu$m,
and [N\,{\sc iii}] at $\lambda=57$ $\mu$m) provide both important diagnostics
of the excitation conditions and a mechanism for cooling. These diagnostic
signatures became widely accessible within the Milky Way, by virtue of the
{\it Kuiper Airborne Observatory} (KAO, see Erickson et al.\ 1984) and the
Long-Wavelength Spectrograph (LWS, Clegg et al.\ 1996) onboard the {\it
Infrared Space Observatory} (ISO, Kessler et al.\ 1996).

The gas-rich dwarf companions to the Milky Way, the Large and Small Magellanic
Clouds (LMC and SMC), offer a unique opportunity for a global assessment of
the feedback into the ISM and the conditions for star formation, something
which is much more challenging to obtain for the Milky Way due to our position
within it. The LMC and SMC are nearby ($d\approx50$ and 60 kpc, respectively:
Cioni et al.\ 2000; Keller \& Wood 2006) and already the scanning survey with
the {\it IR Astronomical Satellite} (IRAS, Neugebauer et al.\ 1984) showed
discrete sources of far-IR emission in them. The star-forming regions, massive
and intermediate-mass stars, and ISM are also lower in metal content than
similar components of the Galactic Disc, $Z_{\rm LMC}\approx0.4$ Z$_\odot$ and
$Z_{\rm SMC}\approx0.1$--0.2 Z$_\odot$ (cf.\ discussion in Maeder, Grebel \&
Mermilliod 1999). This offers the possibility to assess the effect metallicity
has on the dust content and on the heating and cooling processes, and to study
these in environments that are more similar to those prevailing in the early
Universe than the available Galactic examples (cf.\ Oliveira 2009).

The {\it Spitzer Space Telescope} (Werner et al.\ 2004) marries superb
sensitivity with exquisite imaging quality, able to detect the far-IR emission
from a significant fraction of the total populations of YSOs, massive red
supergiants (RSGs), intermediate-mass Asymptotic Giant Branch (AGB) stars,
Planetary Nebulae (PNe), and rare but extreme --- and important --- phases in
the late evolution of massive stars such as Luminous Blue Variables (LBVs) and
SNRs. The telescope also carried a facility, the MIPS-SED, to obtain spectra
at 52--93 $\mu$m, and we used this to target representative samples of
luminous 70-$\mu$m point sources in the LMC and SMC. The SMC spectra are
presented in Paper II in this two-part series (van Loon et al.\ 2009); here we
present the results of the LMC observations.

\section{Observations}

\subsection{Data collection and processing}

\begin{figure*}
\epsscale{1.18}
\plotone{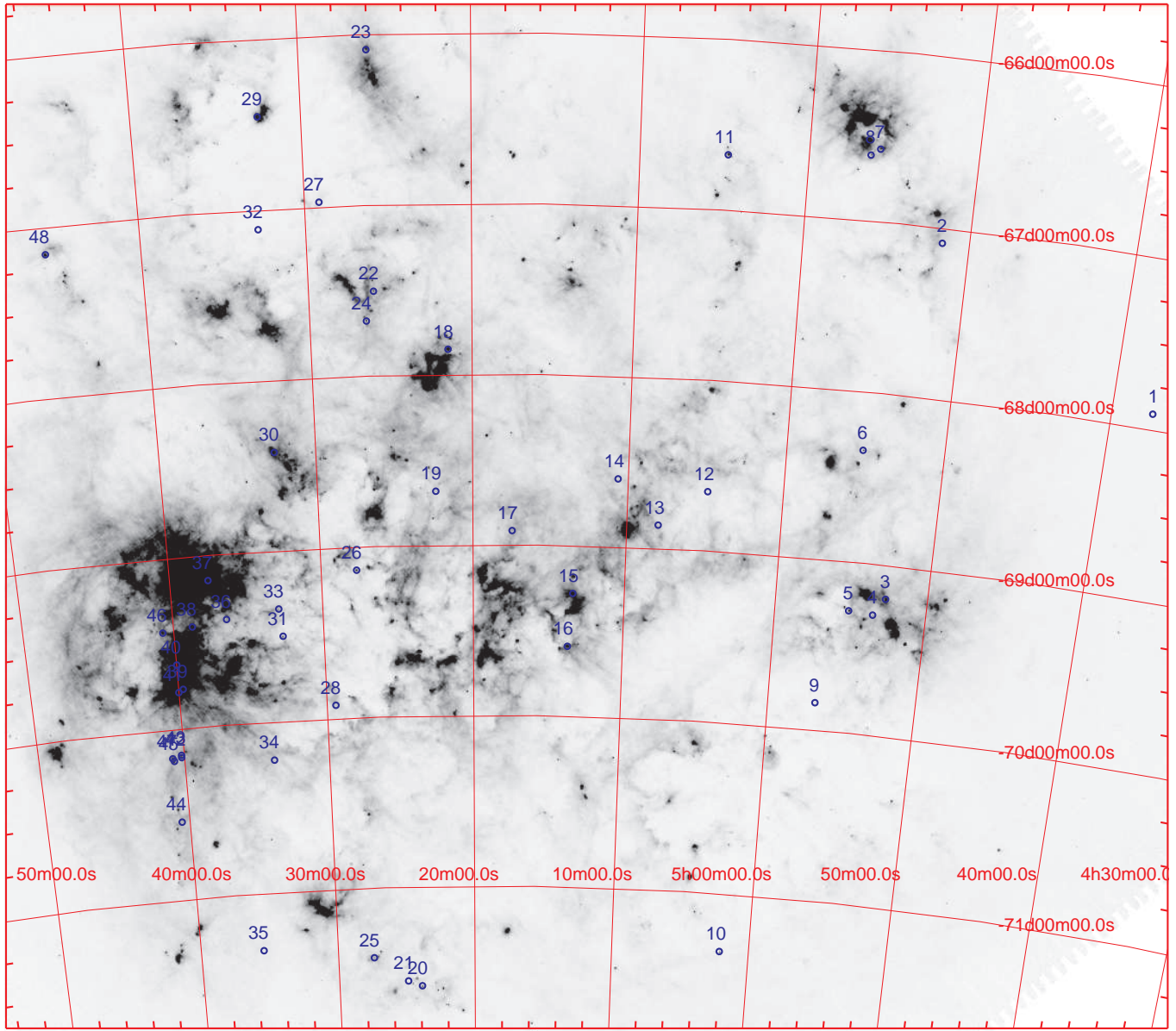}
\caption{All 48 MIPS-SED point sources plotted on top of the MIPS 70-$\mu$m
SAGE-LMC image. The brightest, large region of diffuse 70-$\mu$m emission, the
30\,Doradus mini-starburst dominates the East of the LMC, with the molecular
ridge extending from it to the South. It is separated from other H\,{\sc ii}
regions scattered throughout the LMC by a huge cavity around $5^{\rm h}30^{\rm
m}$, $-69^\circ$, more than a degree across.}
\label{f1}
\end{figure*}

Our dataset comprises low-resolution spectra obtained using the Spectral
Energy Distribution (SED) mode of the {\it Multiband Imaging Photometer for
Spitzer} (MIPS; Rieke et al.\ 2004) onboard the {\it Spitzer Space Telescope}
(Werner et al.\ 2004), taken as part of the SAGE-Spec {\it Spitzer} Legacy
Program (Kemper et al.\ 2009). The spectra cover $\lambda=52$--93 $\mu$m, at a
spectral resolving power $R\equiv\lambda/\Delta\lambda=15$--25 (two pixels)
and a cross-dispersion angular resolution of 13--24$^{\prime\prime}$
Full-Width at Half-Maximum (sampled by $9.8^{\prime\prime}$ pixels). The slit
is $20^{\prime\prime}$ wide and $2.7^\prime$ long, but $0.7^\prime$ at one end
of the slit only covers $\lambda>65$ $\mu$m as a result of a dead readout. To
place the angular scales into perspective, $20^{\prime\prime}\equiv5$ pc at
the distance of the LMC. This is characteristic of a SNR, star cluster, or
molecular cloud core; it is smaller than a typical H\,{\sc ii} region, but
larger than a typical PN.

\begin{figure*}
\epsscale{1.1}
\plotone{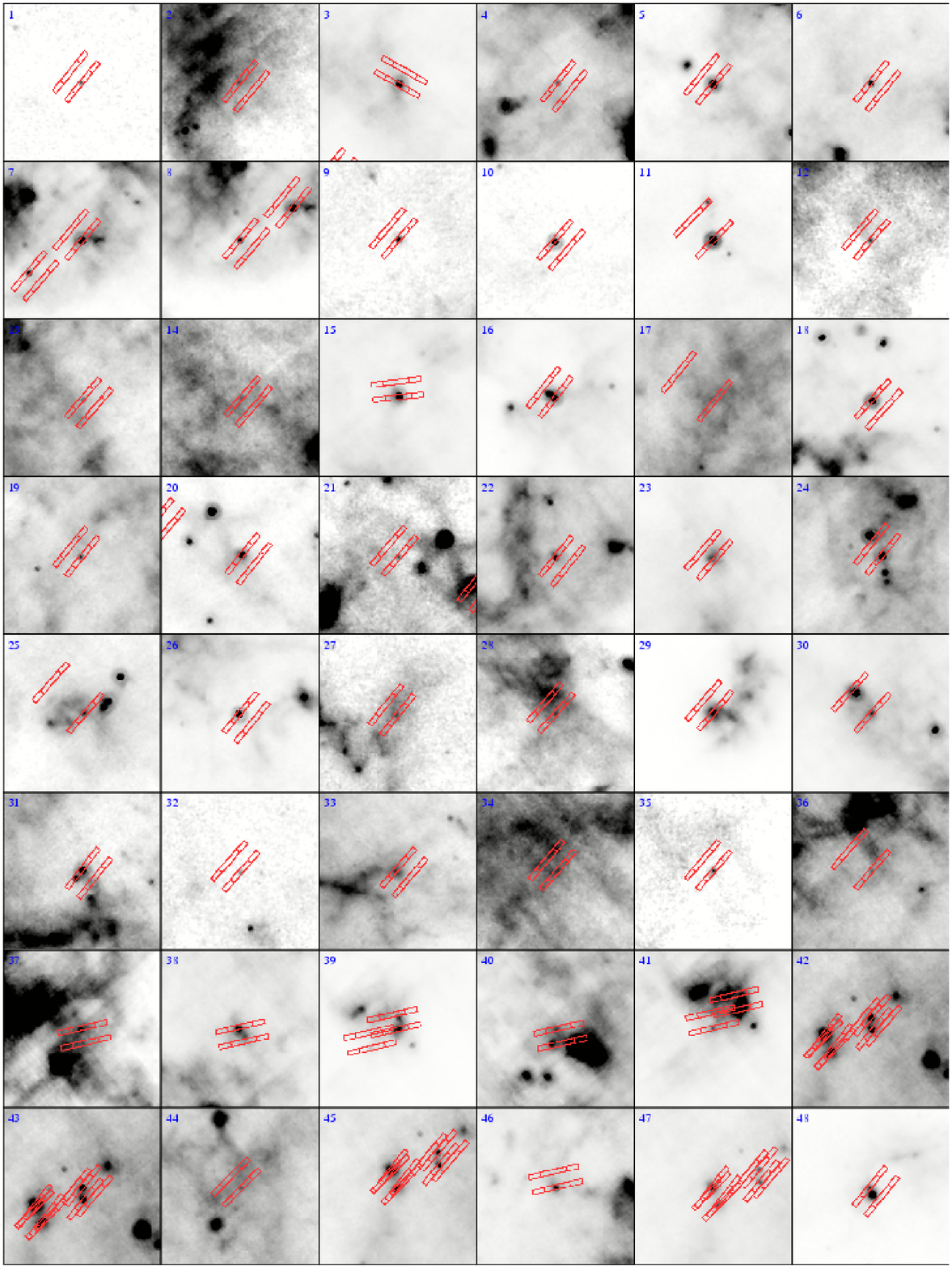}
\caption{Close-ups of the 70-$\mu$m emission centred on each of the 48
MIPS-SED targets, with overplotted the AOR footprints (on- and off-source slit
orientations). All images have North up and East to the left, and measure
$10^\prime$ on each side. The intensity scale is linear, but adjusted
individually such as to facilitate an assessment of the relative brightness of
the target compared to the background.}
\label{f2}
\end{figure*}

The target list, Table 1, is described in \S 2.2 and \S 3, and their
distribution on the sky is displayed in Figure 1. The background spectrum was
measured at one of four possible chop positions, chosen to be free of other
discrete sources of 70-$\mu$m emission. This depends on the time of
observation, and as the observations were scheduled earlier than anticipated,
the off-source position was not always empty. Figure 2 shows 70-$\mu$m
close-ups, extracted from the SAGE-LMC {\it Spitzer} Legacy Program (Meixner
et al.\ 2006), with the Astronomical Observation Request (AOR) footprints
overlain.

The raw data were processed with the standard pipeline version S16.1.1, and
the spectra were extracted and calibrated using the {\sc dat} software, v3.06
(Gordon et al.\ 2005). Spectra were extracted from the on-off
background-subtracted frame, unless the spectrum at the off position was
affected by discrete sources; then the spectrum of the source was extracted
from the on-source observation only. The extraction aperture was five pixels
wide in the cross-dispersion direction, and the (remaining) background level
was determined in a-few-pixel-wide apertures at either side of, and at some
distance from, the extraction aperture. The extracted spectrum was corrected
to an infinite aperture and converted to physical units, providing an absolute
flux calibrated spectrum (cf.\ Lu et al.\ 2008). The spectra extracted from
the on-source frames may be affected by detector artifacts, which are
otherwise cancelled by subtracting the off-source frame. The quality of the
spectrum extraction (a subjective assessment) is listed in Table 2, along with
other MIPS-SED descriptors.

In our analysis of the MIPS-SED data, we shall also make use of associated
photometry, from SAGE at 24, 70, and 160 $\mu$m with MIPS.

\subsection{Target selection}

\begin{figure}
\epsscale{1.16}
\plotone{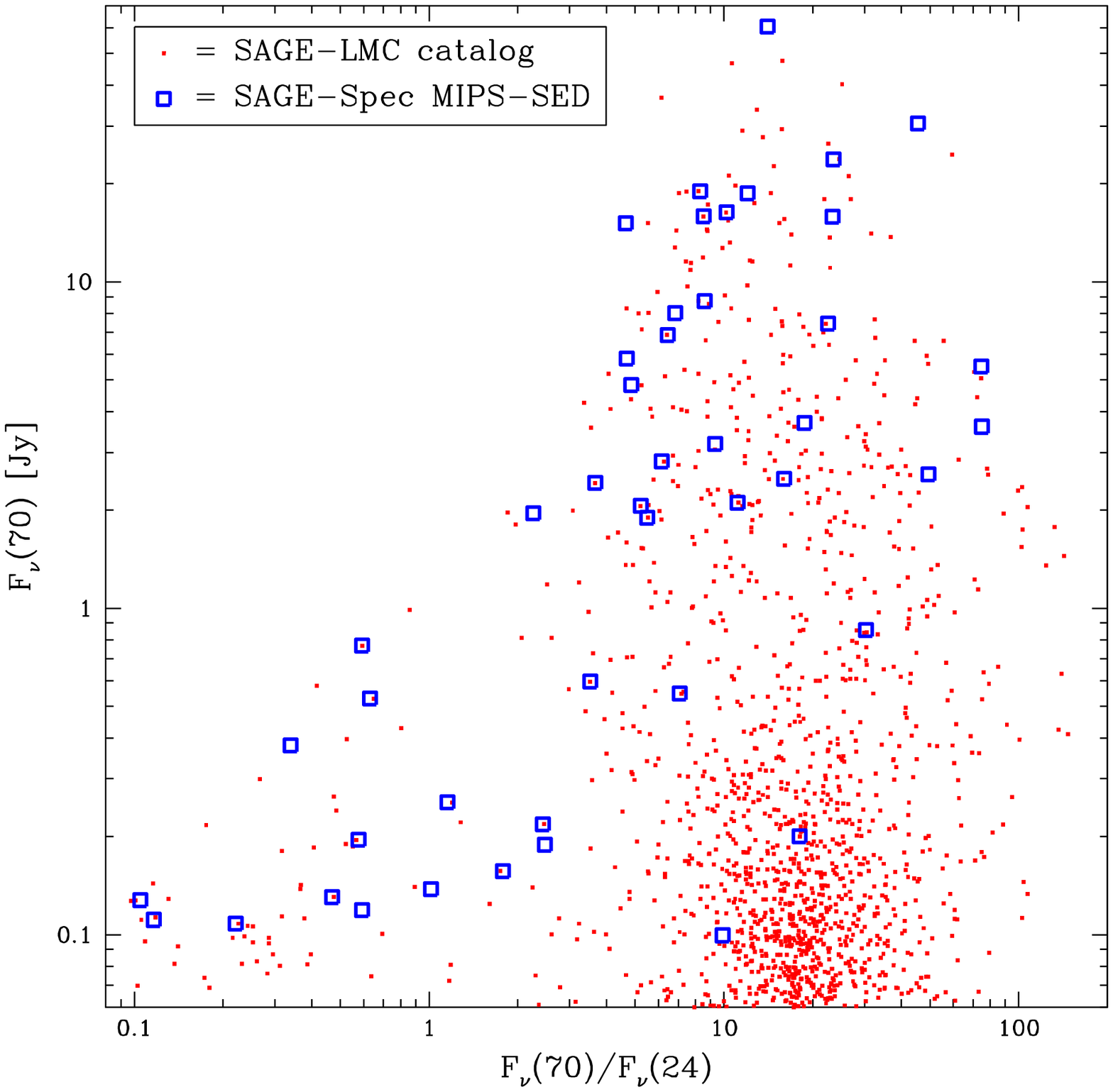}
\caption{$F_\nu(70)$ vs.\ $F_\nu(70)/F_\nu(24)$ diagram, with MIPS photometry
from the SAGE-LMC catalog (dots) and for the SAGE-Spec MIPS-SED targets
(squares). For 70 $\mu$m it is the mosaicked photometry extraction and for the
24 $\mu$m photometry it is epoch-1 only.}
\label{f3}
\end{figure}

The targets were selected on the basis of the following criteria: (i) point
source appearance at 70 $\mu$m, and (ii) a minimum flux density at 70 $\mu$m
of $F_\nu(70)>0.1$ Jy (for MIPS-SED detection purposes). Further criteria were
applied to reduce the large sample of potential targets to within a reasonable
time request: (iii) requirement at the time of proposal submission to have a
{\it Spitzer} IRS spectrum in the archive or be a target for a planned IRS
observation, (iv) aim to cover the F$_\nu(70)$ vs.\ F$_\nu(70)$/F$_\nu(24)$
diagram uniformly, and (v) aim to maximize the diversity of objects as far as
identifications existed. Table 3 summarizes the MIPS photometric properties of
the selected targets, and Figure 3 shows them in the F$_\nu(70)$ vs.\
F$_\nu(70)$/F$_\nu(24)$ diagram in comparison to the entire population of
70-$\mu$m point sources from SAGE-LMC.

The variability at 24 $\mu$m is defined here as:
\begin{equation}
Var \equiv 100\times\left(F_{\nu,\ {\rm max}}/F_{\nu,\ {\rm min}}-1\right),
\end{equation}
and listed in Table 3. There is a fair correlation between this index and
known large-amplitude, long-period variability. For example, MSX-LMC\,349 is a
carbon star and IRAS\,05298$-$6957 is an OH/IR star, both showing the
variability expected in these evolutionary stages and indeed observed for
these sources at shorter wavelengths (see \S 3). In some cases, the index may
be affected by source confusion, and it is helpful in interpreting the
photometry and its reliability, and the nature of the MIPS-SED target.
Examples where this is likely to have happened are N\,159-P2, UFO\,1 and
MSX-LMC\,956. An interesting large-amplitude variable source is the YSO
30\,Dor-17, showing a variation by a factor 1.6.

The targets cover a range of object types as well as IR luminosities, and they
are well-spread across the LMC galaxy (Figs.\ 1 \& 3). The targets are
affected by high, complex background emission and source confusion to a
varying degree; some sit well-isolated in an IR-quiet patch whilst others ---
even though bright --- can be difficult to extract from surrounding diffuse
emission (Fig.\ 2). Although there is a slight tendency towards luminous YSOs
and compact H\,{\sc ii} regions in the prominent star-forming regions of the
LMC, this bias is rather mild. In fact, many of the most luminous point
sources were not observed, and some well-known objects did not make it to the
selection, for instance the LBVs S\,Dor and R\,127. But several objects are
observed, which are closely grouped together in the molecular ridge South of
the 30\,Doradus mini-starburst region. This might allow to catch a glimpse of
evolutionary hierarchy.

\section{Comments on individual objects}

In the remainder of this paper, we shall refer to objects from the Henize
(1956) catalog as ``N\,[number]''; the full designation would be
``LHA\,120-N\,[number]''. Likewise, objects from the Reid \& Parker (2006)
catalog shall be referred to as ``RP\,[number]''. Three sources with only a
SAGE designation are abbreviated following the IRAS convention (where the last
digit of the RA part derives from decimal minutes). Table 1 describes all
MIPS-SED targets, with literature references checked until Spring 2009.

Most targets have been studied before, and brief summaries of their nature are
given below. However, next to nothing is known about SAGE\,04374$-$6754,
SAGE\,05223$-$6841, MSX-LMC\,577, IRAS\,05281$-$7126, MSX-LMC\,741, UFO\,1 (we
note here, from inspection of ESO B-band images, a coincidence with a faint
optical point source very close to another bright point source), RP\,85, and
SAGE\,05407$-$7011.

\subsection{SMP-LMC\,11 (\#2)}

Discovered as a PN candidate by Sanduleak, MacConnell \& Philip (1978), it was
classified as an extreme AGB star by Blum et al.\ (2006). The central star was
not detected (Villaver, Stanghellini \& Shaw 2007). Shaw et al.\ (2006) found
it to be the most compact bipolar PN in the LMC, with an outer arc which they
speculate could be part of a faint halo (or a bow-shock?). Dopita, Ford \&
Webster (1985) and Dopita et al.\ (1988) already noted the complex and
energetic internal dynamics of the ionized nebula. Leisy \& Dennefeld (2006)
classified its optical line emission spectrum as Type ``i'': nitrogen-rich
(compared to oxygen), helium-poor, metal-poor. SMP-LMC\,11 has the lowest
argon abundance of all their LMC objects, 1.2 dex below the LMC average, and
only one of their SMC objects is (marginally) more metal-poor. The oxygen
abundance is equally low, which was already noted by Morgan (1984) who found
it low for its (high) excitation class. It has likely a low-mass progenitor,
in line with the upper limit on the ionized mass in the nebula of $<0.21$
M$_\odot$ (Wood et al.\ 1987). Zijlstra et al.\ (1994) identified it with an
IRAS source. Bernard-Salas et al.\ (2006) presented the {\it Spitzer} IRS
spectrum, which shows acetylene, polyacetylenic chains, and benzene, all in
absorption against a 330 K dust emission continuum, but no Polycyclic Aromatic
Hydrocarbons (PAHs) nor nitrogen-bearing molecules.

\subsection{IRAS\,04530$-$6916 (\#3)}

This was first thought to be a RSG. Wood et al.\ (1992) suggested
small-amplitude variability with a period of $P\approx1260$ d, but this was
not confirmed (Whitelock et al.\ 2003). The ISO (12--60 $\mu$m) colors clearly
pointed at oxygen-rich dust (Trams et al.\ 1999). Van Loon et al.\ (1999b)
estimated a very high mass-loss rate on the premise that it is a
dust-enshrouded RSG. Being a very bright far-IR source, van Loon et al.\
(2001b) were subsequently surprised that no maser emission could be
detected, and on inspection of the IRAS color--color diagram suggested that it
could be a YSO. Van Loon et al.\ (2005a) provided evidence in support of this
assertion, in the form of an optical spectrum showing a B[e]-type
emission-line spectrum on top of a composite of a warm and cold continuum.
Sloan et al.\ (2008) presented the {\it Spitzer} IRS spectrum, which displays
Class A PAH emission pointing also at a star-forming nature, and silicate
absorption against a cool dust emission continuum. It excites the nebula
N\,81A (Henize 1956) = DEM\,L\,15a (Davies, Elliott \& Meaburn 1976). It was
detected in the radio survey of Filipovi\'c et al.\ (1995), as source
LMC\,B0453$-$6917. The nebula's radio and IR properties are commensurate with
it being an H\,{\sc ii} region (Filipovi\'c et al.\ 1998c; see also Voges et
al.\ 2008).

\subsection{IRAS\,04537$-$6922 (\#4)}

The exciting star of N\,82 (Henize 1956), Brey\,3a was suggested by
Heydari-Malayeri et al.\ (1990) to be a WC9-type Wolf-Rayet (WR) star, the
first to have been discovered in the Magellanic Clouds. They detected C\,{\sc
iii} lines and found it nitrogen-rich and oxygen-poor. They also found
evidence for another component, possibly a main-sequence O8-type companion.
Breysacher, Azzopardi \& Testor (1999) still listed this as its likely nature,
even though Moffat (1991) had already argued that it is not a WC9 star but
more likely a WNL or Of star. This was confirmed by Heydari-Malayeri \&
Melnick (1992), who noted the narrow C\,{\sc iv} line (200--300 km s$^{-1}$)
and suggested it may be a transition object between an Of or Of?p star and a
WR star. It was already discovered as an emission-line star by Lindsay \&
Mullan (1963), LM\,1-6; Meynadier \& Heydari-Malayeri (2007) included it in
their new class of low-excitation blobs. We note the similarity with
SMC\,LMC\,11. Egan, van Dyk \& Price (2001) associated the IR source with the
M3-type red giant GV\,60, which is only $5.2^{\prime\prime}$ away from
Brey\,3a. This star had been listed as an M-type supergiant in Westerlund,
Olander \& Hedin (1981), as WOH\,S\,60. Although emission in the {\it Spitzer}
IRAC, IRS and MIPS 24-$\mu$m observations might have a contribution from this
cool giant, there is no indication that it should be bright in the MIPS-SED
range and in the MIPS 70-$\mu$m filter. We therefore consider Brey\,3a the
source of the far-IR emission.

\subsection{N\,89 (\#5)}

The emission-line star LM\,1-8 (Lindsay \& Mullan 1963) was recognized as a
very low-excitation nebula by Sanduleak \& Philip (1977). An association with
a nearby IRAS source was made in Loup et al.\ (1997). Considering it a compact
H\,{\sc ii} region, Joblin et al.\ (2008) discussed the {\it Spitzer} IRS
spectrum of this object, showing evidence for both neutral and
positively-charged PAHs (cations) as well as a population of very small
grains. This suggests a photon-dominated region, either associated with an
irradiated dust shell of an evolved object or with an H\,{\sc ii}--molecular
cloud interface in a star-forming region.

\subsection{WOH\,G064 (\#6)}

WOH\,G064 was discovered by Westerlund et al.\ (1981) as an unremarkable red
giant. It was soon found out by Elias, Frogel \& Schwering (1986) to be a very
luminous RSG; their mid-IR photometry revealed a thick dust shell with the
10-$\mu$m silicate dust feature in self-absorption. They classified the
optical spectrum as M7.5. Forbidden-line emission and at least some of the
H$\alpha$ emission were shown by van Loon et al.\ (2001b) to arise in the ISM.
The self-absorbed silicate feature was confirmed in a groundbased 8--13 $\mu$m
spectrum (Roche, Aitken \& Smith 1993), IRAS LRS spectrum (Kwok, Volk \&
Bidelman 1997), ISOPHOT-S spectrum (Trams et al.\ 1999), and {\it Spitzer} IRS
spectrum (Buchanan et al.\ 2006). Van Loon et al.\ (1999b, 2005a) estimated a
mass-loss rate, $\dot{M}\sim10^{-3}$ M$_\odot$ yr$^{-1}$, and luminosity,
$L\approx4$--$5\times10^5$ L$_\odot$. Ohnaka et al.\ (2008) resolved the dust
shell using the VLTI, proving the suspected bipolar geometry, leading to a
down-ward revision of the luminosity, $L\approx2.8\times10^5$ L$_\odot$. They
derived a circumstellar envelope mass of 3--9 M$_\odot$, for a gas-to-dust
ratio $\psi=200$--500. WOH\,G064 varies with an amplitude $\Delta K\approx0.3$
mag and a period $P=841$ (Whitelock et al.\ 2003) -- 930 d (Wood et al.\
1992).

Wood, Bessell \& Whiteoak (1986) detected 1612 MHz OH maser emission from the
wind; Wood et al.\ (1992) also presented the detection of one of the main
lines at 1665 MHz. Van Loon et al.\ (1996) detected 86 GHz SiO maser emission,
challenging the interpretation of the OH maser profile and the low wind speed
derived from it. Van Loon et al.\ (1998b) detected 22 GHz H$_2$O maser
emission coincident with the SiO maser peak. They also presented an echelle
spectrum of the Ca\,{\sc ii} IR triplet showing light scattered off the
expanding circumstellar dust envelope. Van Loon et al.\ (2001b) presented an
improved SiO maser emission profile, possibly revealing absorption of the
receding part, as well as an improved H$_2$O maser profile showing also the
satellite peaks indicating an accelerating wind. Finally, Marshall et al.\
(2004) detected the red peak of the OH maser emission, confirming the revised
wind speed of $v_{\rm wind}\approx25$ km s$^{-1}$.

\subsection{IRAS\,04557$-$6639 (\#7)}

This object is situated at the South-Western rim of the bubble created by the
OB association LH\,9 that gives the second largest star-forming complex in the
LMC, N\,11, its characteristic shape. It is embedded in the molecular cloud
N11-03 (Caldwell \& Kutner 1996; Israel et al.\ 2003) and the H\,{\sc ii}
region N\,11I (Henize 1956). Rosado et al.\ (1996) noted that the embedded
star in N\,11I is unknown --- we may now have found it. We identify it with
the Herbig Ae/Be star \#25 in Hatano et al.\ (2006), with $JHK_{\rm
s}=16.51\pm0.06,15.84\pm0.07,13.97\pm0.07$ mag.

\subsection{IRAS\,04562$-$6641 (\#8)}

On the Southern rim of N\,11 (Henize 1956), and close to the molecular cloud
N\,11-02 (Caldwell \& Kutner 1996; Israel et al.\ 2003), we identify this
source with the Herbig Ae/Be star \#44 in Hatano et al.\ (2006), with
$JHK_{\rm s}=16.04\pm0.14,15.03\pm0.04,14.24\pm0.10$ mag.

\subsection{R\,66 (\#9)}

S\,73 in Henize (1956), better known as R\,66, was typified by Feast,
Thackeray \& Wesselink (1960) as Aeq, where ``e'' stands for emission lines
and ``q'' for P\,Cygni-type line profiles. It is now classified as a B8\,Ia
supergiant showing the B[e] phenomenon (Lamers et al.\ 1998b). Stahl et al.\
(1983) derived a temperature $T_{\rm eff}=12,000$ K, luminosity
$L=3\times10^5$ L$_\odot$, stellar mass $M_{\rm ZAMS}=30$ M$_\odot$, and
mass-loss rate $\dot{M}\approx3\times10^{-5}$ M$_\odot$ yr$^{-1}$. It displays
photometric $\alpha$\,Cyg-type variability, and may be a precursor of an
S\,Dor-type LBV (van Genderen \& Sterken 2002). It is also variable at near-IR
wavelengths, $\Delta K\approx0.1$ mag (Wood et al.\ 1992). It is below the
Humphreys-Davidson limit and rather cool (Zickgraf et al.\ 1986), and we thus
suggest that it may be a blue-loop star on its way back to becoming a RSG.
Nebular lines of low ionization energy indicated a PN-like shell (Stahl \&
Wolf 1986). Modelling of Fe\,{\sc ii} lines suggested that apart from the 300
km s$^{-1}$ wind there is a disc (Muratorio \& Friedjung 1988); this was
corroborated by CO $1^{\rm st}$-overtone emission at 2.3 $\mu$m (McGregor,
Hyland \& Hillier 1988) and polarization measurements (Magalh\~aes 1992).
Kastner et al.\ (2006) presented the {\it Spitzer} IRS spectrum showing a flat
continuum with silicate emission, including crystalline material as well as
(weak) PAHs. They point out the similarity with Herbig Ae/Be discs. Given the
composition this may be a debris disc, although it would have had to survive
for $\sim10$ Myr.

\subsection{R\,71 (\#10)}

S\,155 in Henize (1956), better known as R\,71, is one of the few LBVs in the
LMC (for a review on LBVs see Humphreys \& Davidson 1994). Feast et al.\
(1960) determined a spectral type B2.5\,Iep at minimum light, whilst Thackeray
(1974) determined a much cooler type of B9ep--A1eq at maximum light; a similar
spectral change was observed from 2006 to 2008 (Munari et al.\ 2009). Lennon
et al.\ (1993) determined a spectroscopic mass $M_{\rm today}=20\pm2$
M$_\odot$, and an initial mass $M_{\rm ZAMS}\approx40$--45 M$_\odot$ based on
the temperature $T_{\rm eff}=17,000$--17,500 K and luminosity $\log L/{\rm
L}_\odot=5.85\pm0.04$. They note the small oxygen-to-nitrogen ratio, implying
strong CNO processing. It is in the S\,Dor instability strip (e.g., Smith,
Vink \& de Koter 2004), possibly post-RSG, and expected to explode as a
supernova type Ib/c (Smith \& Conti 2008). Van Genderen (1989) shows that
R\,71, like other S\,Dor variables can be distinguished from $\alpha$-Cyg-type
variables by the higher amplitude of its micro-variability. Weak [N\,{\sc ii}]
nebular emission was detected by Stahl \& Wolf (1986) who also measured a wind
velocity of $\approx158$ km s$^{-1}$ from the H$\alpha$ line profile. Glass
(1984) detected 10-$\mu$m emission from circumstellar dust. Wolf \& Zickgraf
(1986) then identified it with an IRAS source. Roche et al.\ (1993) detected
the 10-$\mu$m silicate feature in emission, and surmised that the
longer-wavelength portion of the SED implies expulsion of significant amounts
of dust $>10^4$ yr ago. Voors et al.\ (1999) detected crystalline silicates
and PAHs in an ISOCAM+SWS spectrum, and derived a dust mass of $M_{\rm
dust,\ recent}\approx0.02$ M$_\odot$, ejected at a rate
$\dot{M}\geq7\times10^{-4}$ M$_\odot$ yr$^{-1}$ over the past 3000 yr. They
noted that the far-IR emission implies $M_{\rm dust,\ ancient}\approx0.3$
M$_\odot$ of additional dust ejected before that. They ascribed also the inner
dust shell to previous RSG mass loss, but we note that it takes $>10^4$ yr to
evolve from the RSG to where we find R\,71 today (van Genderen 2001). The
presence of PAHs in the oxygen-rich wind may indicate non-equilibrium
chemistry. That R\,71 has already lost a significant fraction of its mass is
further corroborated by the large $Q$-values of its pulsations (Lamers et al.\
1998b).

\subsection{IRAS\,05047$-$6644 (\#11)}

We identify this IR source with the PN candidate RP\,1933. Although centred
$2^{\prime\prime}$ away from the IR position, the PN has a diameter of
$18^{\prime\prime}$ (Reid \& Parker 2006) and thus easily encompasses the IR
source. Buchanan et al.\ (2006) presented the {\it Spitzer} IRS spectrum,
displaying strong PAHs on a rising continuum; they estimated an IR luminosity
of $L_{\rm IR}\approx6.8\times10^4$ L$_\odot$. This appears to make the object
more luminous than the classical AGB limit, and casts doubt on a
classification as a canonical PN. It must have a more massive progenitor.

\subsection{SMP-LMC\,21 (\#12)}

HST observations (Stanghellini et al.\ 1999; Vassiliadis et al.\ 1998b) of
this high-excitation PN (Morgan 1984) revealed a
$1.15^{\prime\prime}$-diameter quadrupolar nebula, with an expansion velocity
of 27 km s$^{-1}$ --- we note that Dopita et al.\ (1988) measured $\approx50$
km s$^{-1}$. Reid \& Parker (2006) listed a much larger $14^{\prime\prime}$
diameter, so it may have an extended halo. The PN is oxygen-poor compared to
nitrogen as well as argon, which suggests a relatively massive AGB progenitor,
normal for Type I PNe (Leisy \& Dennefeld 2006; cf.\ Dufour \& Killen 1977;
Leisy \& Dennefeld 1996 --- who also noted a strong, red continuum in the
optical spectrum). Pe\~na et al.\ (1994) claimed the central star is a WC-type
WR star, at odds with a relatively massive AGB progenitor. Barlow (1987), Wood
et al.\ (1987) and Boffi \& Stanghellini (1994) determined an ionized mass in
the nebula of 0.6, $<0.3$ and 0.243 M$_\odot$, respectively. Stanghellini et
al.\ (2007) presented the {\it Spitzer} IRS spectrum showing crystalline
silicates and both low- and high-excitation nebular emission lines, on a 130 K
dust continuum. It has by far the strongest IR excess of the PNe in their
sample.

\subsection{SMP-LMC\,28 (\#13)}

This is a compact PN ($0.58^{\prime\prime}\times0.35^{\prime\prime}$) composed
of three emission blobs, one is centred on the central star and the other two
are faint arms particularly bright in [N\,{\sc ii}] (Shaw et al.\ 2001). The
PN expands at 55 km s$^{-1}$ (Dopita et al.\ 1988) and is particularly rich in
nitrogen (Stanghellini et al.\ 2005). Villaver, Stanghellini \& Shaw (2003)
derived a luminosity, $\log L/{\rm L}_\odot=4.14\pm0.05$, and temperature,
$T<1.7\times10^4$ K, which would suggest a $\approx3$ M$_\odot$ AGB
progenitor. The ionized mass in the nebula is $<0.19$ M$_\odot$ (Wood et al.\
1987). Zijlstra et al.\ (1994) identified the PN with IRAS\,05081$-$6855. The
{\it Spitzer} IRS spectrum (Bernard-Salas et al.\ 2009) shows a high neon
abundance (Bernard-Salas et al.\ 2008).

\subsection{SMP-LMC\,36 (\#14)}

Van Loon et al.\ (2006) noticed the proximity of the PN to the mid-IR source
MSX-LMC\,45 = IRAS\,05108$-$6839, but rule out that they are the same: the
PN is $9^{\prime\prime}$ distant from a very red near-IR point source that
dominates also at mid-IR wavelengths. Classified as an OH/IR star by Egan et
al.\ (2001), the IRAS/MSX source was identified on the basis of a 2.8--4.1
$\mu$m spectrum by van Loon et al.\ (2006) as a luminous, $\log L/{\rm
L}_\odot=4.31$, dust-enshrouded carbon star with an estimated mass-loss rate
$\dot{M}\approx5\times10^{-5}$ M$_\odot$ yr$^{-1}$. Although it is conceivable
that the carbon star has a cold dust shell, we expect that the PN will
dominate at far-IR wavelengths (Hora et al.\ 2008). The {\it Spitzer} IRS
spectrum is dominated by emission lines and some PAHs, and thus clearly that
of the PN (Bernard-Salas et al.\ 2008). The PN has unremarkable abundances,
except for a high helium abundance of $N({\rm He})/N({\rm H}=0.142$ (de
Freitas Pacheco, Costa \& Maciel 1993; cf.\ Henry 1990). An expansion velocity
of 35 km s$^{-1}$ was measured by Dopita et al.\ (1988).

\subsection{IRAS\,05137$-$6914 (\#15)}

Coinciding with the H\,{\sc ii} object N\,112 (Henize 1956) = DEM\,L\,109
(Davies et al.\ 1976), powered by the star cluster OGLE-CL\,LMC241
(Pietrzy\'nski et al.\ 1999), the mid-IR source is associated with a bright,
unresolved radio continuum source, and is likely a compact H\,{\sc ii} region
(Mathewson et al.\ 1985). Boji\v{c}i\'c et al.\ (2007) carried out a detailed
analysis of radio and optical data, which beautifully show its location at the
NE edge of the SNR\,B0513$-$692 (Mathewson et al.\ 1985). They found a second
SNR\,J051327$-$6911 due SE from the IRAS source. Although they did not see any
evidence for interaction between the two SNRs and either of the SNRs and the
compact H\,{\sc ii} region they are all clearly associated with the same
star-forming region. A molecular cloud was detected with an unusually narrow
CO profile, $\Delta v\approx3$ km s$^{-1}$ (Israel et al.\ 1993).

\subsection{MSX-LMC\,222 (\#16)}

Buchanan et al.\ (2006) presented a {\it Spitzer} IRS spectrum, which they
argued lacks the high-ionization lines typical of PNe. This, and the extended
mid-IR emission seen in the MSX band A image, they argued suggests that
MSX-LMC\,222 is associated with star formation. The nearby
($16^{\prime\prime}$), possibly associated cold IRAS source,
IRAS\,05141$-$6938, was searched for methanol emission by Beasley et al.\
(1996); no such emission was detected.

\subsection{MSX-LMC\,349 (\#17)}

Egan et al.\ (2001) classified this red source as an OH/IR star, but a
3-$\mu$m spectrum revealed it to be a dust-enshrouded carbon star (van Loon et
al.\ 2006). This was confirmed with a {\it Spitzer} IRS spectrum showing
strong SiC emission at 11.3 $\mu$m and MgS emission at $\lambda\approx30$
$\mu$m (Zijlstra et al.\ 2006). Groenewegen et al.\ (2007) derived a
luminosity of $L=7700$ L$_\odot$ and a mass-loss rate of
$\dot{M}=8\times10^{-6}$ M$_\odot$ yr$^{-1}$; they also determined a period of
$P=600$ d for its variability.

\subsection{IRAS\,05216$-$6753 (\#18)}

The nature of the bright H$\alpha$ knot N\,44A (Henize 1956) is elusive. Reid
\& Parker (2006) classified it as a ``true'' PN, albeit with a rather large
H$\alpha$ diameter of $15.7^{\prime\prime}$. It is a very luminous IRAS
source, though, and unlikely to be a PN descending from an AGB star. Roche et
al.\ (1987) detected emission from silicate dust in a groundbased 10-$\mu$m
spectrum of the associated bright, cool mid-IR source IRAS\,05216$-$6753 =
TRM\,11. The global SED and lack of conspicuous variability bear resemblance
to IRAS\,04530$-$6910 (Wood et al.\ 1992). Indeed, in both cases van Loon et
al.\ (2001b) suspected it is a dust-enshrouded, but hot, massive star (cf.\
Zijlstra et al.\ 1996). Chen et al.\ (2009) performed an in-depth assessment
of all available data and concluded that the ionizing power requires an O9\,I
star, but that it is unclear whether it is young or evolved.

\subsection{HS\,270-IR1 (\#20)}

Egan et al.\ (2001) classified this cool mid-IR source as an OH/IR star. Van
Loon, Marshall \& Zijlstra (2005) subsequently associated it with a heavily
reddened near-IR point source in the star cluster HS\,270 (Hodge \& Sexton
1966); from modelling the SED they inferred a possible nature as a post-AGB
star. However, CO$_2$ ice was tentatively detected in the {\it Spitzer} IRS
spectrum (Oliveira et al.\ 2009) and we thus reclassify it as a YSO.

\subsection{SMP-LMC\,62 (\#21)}

IRAS\,05257$-$7135 is associated with the emission-line object N\,201 (Henize
1956; Loup et al.\ 1997; Leisy et al.\ 1997). This is a high-excitation PN,
SMP-LMC\,62 (Westerlund \& Smith 1964; Sanduleak et al.\ 1978; Morgan 1984).
Hora et al.\ (2008) presented the {\it Spitzer} SED, Bernard-Salas et al.\
(2009) the {\it Spitzer} IRS spectrum, and Vassiliadis et al.\ (1998b) an HST
image showing a highly flattened ellipsoidal ring with a diameter
$\approx0.5^{\prime\prime}$ (Villaver et al.\ 2007). The PN is oxygen-rich;
the ionized mass is 0.44--0.59 M$_\odot$ (Barlow 1987; Aller et al.\ 1987;
Dopita \& Meatheringham 1991; Boffi \& Stanghellini 1994). From the FUSE
spectrum, Herald \& Bianchi (2004) derived $T_{\rm eff}\approx45,000$ K,
$L=5370$ L$_\odot$, $M_\star=0.65$ M$_\odot$, mass-loss rate
$\dot{M}\sim10^{-8}$ M$_\odot$ yr$^{-1}$, and wind speed $v_\infty\sim1000$ km
s$^{-1}$. Dopita et al.\ (1988) measured an expansion velocity of the ionized
nebula of $v_{\rm exp}=34.6$ (O\,{\sc iii}) -- 47.5 (O\,{\sc ii}) km s$^{-1}$.
Dufour (1991) explained the high N/C ratio and normal helium abundance with
Hot Bottom Burning (Renzini \& Voli 1981), but Leisy \& Dennefeld (1996)
argued that this is inconsistent with the relatively low N/O ratio.
Vassiliadis et al.\ (1998a) used HST spectra to derive a rather high helium
abundance, and in particular a high Si/C ratio. The latter might be explained
by re-accreted products from circumstellar grain destruction. Webster (1976)
noted a similarity with the optical spectra of dusty symbiotic stars;
Feibelman \& Aller (1987) confirmed this possibility on the basis of the low
C\,{\sc iii}\,$\lambda1909$/Si\,{\sc iii}\,$\lambda1892$ ratio. Herald \&
Bianchi (2004) noted the O\,{\sc vi}\,$\lambda\lambda$1032,1038 nebular
emission lines, a rarity for PNe. Like other LMC PNe in their sample, the FUSE
spectrum of SMP-LMC\,62 showed that hot ($T\approx3000$ K) H$_2$ is present.
They argued that a mixture of photo-excitation and shocks is needed to explain
the spectrum. Uniquely in their sample, the H\,{\sc i} in SMP-LMC\,62 appears
to be located within a volume of similar size to that of the ionized nebula.
Interestingly, this was also one of the first extra-galactic radio PNe
(Filipovi\'c et al.\ 2009); its detection at low frequencies is curious as
this is more typical of optically-thick ionized regions which tend to have
ionized masses $<0.1$ M$_\odot$. Filipovi\'c et al.\ argue for a class of
``Super-PNe'' where the radio emission is determined by environmental effects,
with some similarity to SNRs. In conclusion, SMP-LMC\,62 may not be the
product of (canonical) AGB evolution.

\subsection{N\,51-YSO1 (\#22)}

This IR source was classified as an H\,{\sc ii} source by Egan et al.\ (2001).
It is situated in LH\,54, a rich association with stars as early as O8 (Oey
1996), whose stellar winds may have created superbubble DEM\,L\,192 = N\,51D
(Oey \& Smedley 1998). The well-studied WR binary HD\,36402 is only
$22^{\prime\prime}$ to the North-East. MSX-LMC\,824 is located in between YSO1
and YSO2 discovered by Chu et al.\ (2005) in {\it Spitzer} images. YSO1 is the
brighter of the two, and the target of the MIPS-SED observation. Chu et al.\
modelled the SED, deriving $L=10,500$ L$_\odot$ (equivalent to a B2--3
main-sequence star) and an infall rate of $\dot{M}_{\rm acc}=2\times10^{-4}$
M$_\odot$ yr$^{-1}$; the envelope may be as massive as 700 M$_\odot$. Seale et
al.\ (2009) presented the {\it Spitzer} IRS spectrum, displaying silicate
dust absorption at 10 $\mu$m, weak CO$_2$ ice absorption at 15 $\mu$m, weak
fine-structure emission lines, and very weak PAH emission.

\subsection{N\,49 (\#23)}

N\,49 (Henize 1956) = DEM\,L\,190 (Davies et al.\ 1976) was first recognized
as a SNR by Mathewson \& Healey (1964), on the basis of its radio properties,
and by Westerlund \& Mathewson (1966), on the basis of strong [S\,{\sc ii}]
emission lines indicating shocks. It is listed in the Henry-Draper catalog as
HD\,271255 (see Morel 1984), with spectral type ``N'' (for nebular). Situated
in the $\approx10$ Myr-old association LH\,53, its progenitor mass is likely
$M_{\rm ZAMS}\approx20$ M$_\odot$ (Hill et al.\ 1995). Dynamical age estimates
for the SNR range from 4400 yr (Hughes et al.\ 1998) to 6400 yr (Long, Helfand
\& Grabelsky 1981), consistent with the age of the associated $\gamma$-ray
burster pulsar, of 5000 yr (Rothschild, Kulkarni \& Lingenfelter 1994). The
multi-phased shocked nature of the interaction between the ejecta and the ISM
is shown beautifully in X-ray images (e.g., Park et al.\ 2003), far-UV spectra
(Sankrit, Blair \& Raymond 2004), and [Ne\,{\sc v}] images at 0.34 $\mu$m
(Rakowski, Raymond \& Szentgyorgyi 2007); see also the HST images presented by
Bilikova et al.\ (2007), and the earlier work by Shull et al.\ (1985), for
modelling, and Vancura et al.\ (1992b), for a high-resolution multi-wavelength
synthesis. Brogan et al.\ (2004) detected an OH 1720 MHz maser in the Western
part of N\,49.

Graham et al.\ (1987) explained IR emission from the SNR by collisionally
heated dust of 40 K. Williams et al.\ (2006), based on {\it Spitzer} IRAC and
MIPS 24- and 70-$\mu$m images, argued for a lack of PAHs and very small
grains, possibly due to destruction by far-UV radiation from the shock
precursor. They also described the {\it Spitzer} IRS spectrum, of the bright
SE side of the 17-pc-diameter SNR ring (cf.\ Mathewson et al.\ 1983; Shull
1983), which lacks continuum emission from hot dust; Williams et al.\ thus
argued that most of the mid- and far-IR emission arises from line emission.
Our MIPS-SED observation was taken of the same bright spot. Cold CO (Sorai et
al.\ 2001) emission was detected from --- or just outside of --- the same
region, with a virial mass of the molecular cloud $M_{\rm vir}\sim3\times10^4$
(Banas et al.\ 1997) or $2\times10^5$ M$_\odot$ (Mizuno et al.\ 2001).
Modelling of the X-ray spectrum suggested that the SNR has swept up $\sim200$
M$_\odot$ of ISM (Hughes, Hayashi \& Koyama 1998). Dopita (1976) already
argued that the overabundant oxygen and especially nitrogen in the shocked
regions might be due to the release of volatiles as icy grains are destroyed
--- Dennefeld (1986) also argued in favor of grain destruction.

\subsection{IRAS\,05280$-$6910 (\#26)}

Wood et al.\ (1992) associated this IR source with the cluster NGC\,1984, and
suggested it is a supergiant with a birth mass of $M_{\rm ZAMS}\approx15$--20
M$_\odot$. Van Loon et al.\ (2005b) used high-resolution IR imaging to
identify the stellar counterpart, the dust-enshrouded star NGC\,1984-IR1; they
proved it is not the M1 RSG WOH\,G347, at only a few arcseconds distance ---
this star (NGC\,1984-IR2) is much brighter at near-IR wavelengths than IR1 but
much fainter in the mid-IR --- and neither the PN SMP-LMC\,64, which is almost
an arcminute away. This was confirmed by the different slopes of the 3-$\mu$m
spectra of IR1 and IR2 (van Loon et al.\ 2006). Double-peaked OH 1612 MHz
maser emission was detected, as well as an OH 1665 MHz emission peak at a
velocity outside of the 1612 MHz velocity range (Wood et al.\ 1992) and H$_2$O
22 GHz maser emission within the OH 1612 MHz velocity range (van Loon et al.\
2001b).

\subsection{IRAS\,05291$-$6700 (\#27)}

This red variable star, GRV\,0529$-$6700 was discovered by Glass \& Reid
(1985). Reid, Glass \& Catchpole (1988) determined a period of $P=828$ d, but
this was revised to $P=483$ d by Whitelock et al.\ (2003) and $P=503$ d by
Groenewegen et al.\ (2007) (who flagged the variability as being relatively
regular, and note a long secondary period of 3020 d). Van Loon, Zijlstra \&
Groenewegen (1999) presented a 3-$\mu$m spectrum showing C$_2$H$_2$+HCN
absorption: evidence that it is a carbon star. The {\it Spitzer} IRS spectrum
is noisy and rather featureless (Zijlstra et al.\ 2006), but confirms the
carbon star classification. Groenewegen et al.\ (2007) derived a luminosity of
$L=11,900$ L$_\odot$ and a mass-loss rate of $\dot{M}\approx6.6\times10^{-7}$
M$_\odot$ yr$^{-1}$, but their model fit does not reproduce the sharp upturn
at $\lambda>30$ $\mu$m. It is rather puzzling why this unremarkable carbon
star should be so bright at 70 $\mu$m.

\subsection{IRAS\,05298$-$6957 (\#28)}

The classical example of an oxygen-rich massive AGB star, IRAS\,05298$-$6957
shows the most beautiful double-peaked OH 1612 MHz profile of all Magellanic
OH/IR stars (Wood et al.\ 1992); its large-amplitude variability with a very
long period of $P=1280$ d (Wood et al.\ 1992) is entirely consistent with
that. The 3-$\mu$m spectrum (van Loon et al.\ 1999a) and self-absorbed
silicate feature at 10 $\mu$m in the ISOCAM-CVF spectrum (Trams et al.\ 1999)
confirm the oxygen-rich nature. Van Loon et al.\ (1998) noticed its location
in the cluster HS\,327 (Hodge \& Sexton 1966), which led van Loon et al\
(2001a) to estimate a birth mass of $M_{\rm ZAMS}\approx4$ M$_\odot$. Van Loon
et al.\ (1999b) derived a mass-loss rate of $\dot{M}=2.3\times10^{-4}$
M$_\odot$ yr$^{-1}$.

\subsection{IRAS\,05325$-$6629 (\#29)}

Smith, Beall \& Swain (1990) mistakenly identified this IRAS source with the
High-Mass X-ray Binary LMC-X4, which however is $>5^\prime$ away. Egan et al.\
(2001) classified it as a PN. Indebetouw, Johnson \& Conti (2004), using
high-resolution radio continuum observations, classified it as an
ultra-compact H\,{\sc ii} region within N\,55A (Henize 1956), estimating a
B0\,V central star. Buchanan et al.\ (2006) reached the same conclusion, based
on the environment and on the {\it Spitzer} IRS spectrum --- characterized by
prominent PAH and atomic line emission.

\subsection{IRAS\,05328$-$6827 (\#30)}

Identified with a near-IR counterpart by van Loon (2000), it was first
presumed to be an evolved star; Egan et al.\ 2001 classified it as an OH/IR
star. The 3-$\mu$m and {\it Spitzer} IRS spectra, however, revealed
characteristics of a YSO: water ice at 3 $\mu$m, CO$_2$ ice at 15 $\mu$m,
possibly methanol ice, and strong absorption from silicates at 10 and 20
$\mu$m (van Loon et al.\ 2005c; Oliveira et al.\ 2009). It is located in the
relatively isolated N\,148 (Henize 1956) star-forming region.

\subsection{RP\,775 (\#31)}

Although Reid \& Parker (2006) included it in their list of PNe, they noted it
is half-hidden in a more extended H\,{\sc ii} region. We therefore keep open
the possibility that it is a young, or at least massive, stellar object.

\subsection{IRAS\,05329$-$6708 (\#32)}

Identified as a dust-enshrouded star by Reid, Tinney \& Mould (1990), it was
one of the first Magellanic AGB stars from which OH 1612 MHz maser emission
was detected (Wood et al.\ 1992). It has the usual characteristics of an OH/IR
star, such as a very long period of its large-amplitude variability,
$P=1260$--1295 d (Wood et al.\ 1992; Wood 1998; Whitelock et al.\ 2003), and a
strong 10-$\mu$m silicate feature in (self-)absorption (Groenewegen et al.\
1995; Zijlstra et al.\ 1996; Trams et al.\ 1999; Sloan et al.\ 2008). A
mass-loss rate of $\dot{M}=1.8\times10^{-4}$ M$_\odot$ yr$^{-1}$ was derived
by van Loon et al.\ (1999b).

\subsection{MSX-LMC\,783 (\#33)}

Initially classified as a candidate OH/IR star by Egan et al.\ (2001), a {\it
Spitzer} IRS spectrum clearly revealed it is a carbon star (Leisenring et al.\
2008). We note here that the IRS spectrum displayed a dramatic upturn at
$\lambda>20$ $\mu$m, but there is no obvious reason why this source should be
so bright at 70 $\mu$m.

\subsection{HV\,2671 (\#34)}

Alcock et al.\ (1996, 2001) discovered this variable star to be of the
R\,Coronae\,Borealis type, carbon-rich objects experiencing sudden dimmings
followed by slow recovery, with an atypically high $T_{\rm eff}\approx20,000$
K; the optical spectrum at maximum light displays C\,{\sc ii} emission lines.
De Marco et al.\ (2002) favoured an interpretation as a ``born-again''
post-AGB object, i.e.\ a star which has experienced a late thermal pulse
whilst already on the post-AGB track; they estimated a luminosity of $L=6000$
L$_\odot$.

\subsection{R\,126 (\#36)}

This B[e] star --- S\,127 in Henize (1956) --- has been known to display
Balmer line emission for well over a century (Pickering \& Fleming 1897);
Fe\,{\sc ii} and [Fe\,{\sc ii}] lines have been seen in emission too (e.g.,
Stahl et al.\ 1985). The spectral type of B0.5\,Ia$^+$ corresponds to $T_{\rm
eff}\approx22,500$ K and $L\approx1.2\times10^6$ L$_\odot$ (Zickgraf et al.\
1985; cf.\ Shore \& Sanduleak 1984), making it one of the most luminous stars
known. Zickgraf et al.\ (1985) suggested a birth mass of $M_{\rm ZAMS}=70$--80
M$_\odot$. They detected a stellar wind with $v_\infty=1800$ km s$^{-1}$. Van
Genderen \& Sterken (2002) detected small-amplitude brightness variations, and
suggested a link to S\,Dor-type instability. The star appears to dip in
brightness by 0.2--0.3 visual magnitudes, briefly, in a semi-regular way on a
timescale of about a year (see van Genderen et al.\ 2006). Smith (1957)
noticed the steep Balmer decrement; Allen \& Glass (1976) argued that this is
not caused by reddening (cf.\ Koornneef \& Code 1981), but they did supply
early evidence for circumstellar dust. Roche et al.\ (1993) noticed the
weakness of the 10-$\mu$m silicate emission feature. Kastner et al.\ (2006)
presented the {\it Spitzer} IRS spectrum, which is dominated by emission from
amorphous silicates. They estimated that $M_{\rm dust}\sim3\times10^{-3}$
M$_\odot$ is present within an envelope of radius 2500 AU, illuminated by only
15\% of the stellar luminosity. Detailed models including an equatorial disc
were presented by Porter (2003) and Kraus, Borges Fernandes \& de Ara\'ujo
(2007) (cf.\ Zsarg\'o, Hillier \& Georgiev 2008).

\subsection{30\,Dor-17 (\#37)}

Despite its location in the 30\,Doradus mini-starburst region, in the N\,157B
nebula (Henize 1956), this is a little-documented molecular cloud with a
virial mass $M_{\rm vir}\sim1.5\times10^4$ M$_\odot$ (Johansson et al.\ 1998).
It harbors an embedded YSO (Blum et al.\ 2006) and CO$_2$ ice was detected in
the {\it Spitzer} IRS spectrum (Oliveira et al.\ 2009). At $8^{\prime\prime}$
at either side are an emission-line object (Reid \& Parker 2006) and an O7\,V
star (Schild \& Testor 1992); cf.\ Oliveira et al.\ (2006) for IR and 22 GHz
observations of N\,157B.

\subsection{N\,158B (\#38)}

The {\it Spitzer} IRS spectrum (Buchanan et al.\ 2006) of this bright knot in
the N\,158B nebula (Henize 1956) is that of a typical H\,{\sc ii} region, with
emission from PAHs, atomic lines, and dust. It is likely to harbor a massive
star or stellar aggregate, but its age is uncertain.

\subsection{N\,159-P2 (\#39)}

IRAS\,05401$-$6947, in N\,159A (Henize 1956), was resolved into two components
using the ISOCAM instrument, viz.\ LI-LMC\,1501E and W (Comer\'on \& Claes
1998). The Eastern component had been identified by Jones et al.\ (1986) as
the second extra-galactic protostar, P2, using Kuiper Airborne Observatory 50-
and 100-$\mu$m and groundbased near-IR data. It was characterized on the basis
of {\it Spitzer} IRAC images as a Class I protostar of moderate luminosity,
$L\approx4000$ L$_\odot$ (Jones et al.\ 2005). Jones et al.\ refuted an
association between the protostar and the radio continuum emission, which
instead they attributed to the nearby O5 and O7 stars (Deharveng \& Caplan
1992). Nakajima et al.\ (2005) identified a near-IR stellar counterpart, which
was subsequently resolved by Testor et al.\ (2006) into two equally bright and
very red ($(J-K)=4.19$--4.77 mag) stars $0.57^{\prime\prime}$ apart.

\subsection{N\,160-1 (\#40)}

This far-IR source is close to a molecular cloud with an uncertain virial
mass, $M_{\rm vir}\sim7000$ M$_\odot$ (Johansson et al.\ 1998) or $M_{\rm
vir}\sim4.8\times10^4$ M$_\odot$ (Indebetouw et al.\ 2008). One may suspect it
is a YSO, but the association between the IR source and CO emission is not
certain; cf.\ Oliveira et al.\ (2006) for 22 GHz observations of N\,160A.

\subsection{N\,159S (\#41)}

Located in the molecular ridge due South of the 30\,Dor complex, N\,159S is
the brightest knot in the ring-shaped nebula N\,159E (Henize 1956; Israel \&
Koornneef 1991). On {\it Spitzer} IRAC images, N\,159S appears very compact
but noticeably extended (Jones et al.\ 2005). It is near a quiescent molecular
cloud, which has a temperature $T\approx10$ K and density $n\sim10^5$
cm$^{-3}$ (Heikkil\"a, Johansson \& Olofsson 1998). This was confirmed by
Bolatto et al.\ (2000) in spite of it being the brightest [C\,{\sc i}] source
in the region. A large virial mass of $M_{\rm vir}\sim1.5\times10^5$ M$_\odot$
was estimated by Minamidani et al.\ (2008).

\subsection{WOH\,G457 (\#43)}

Although identified with the AGB variable star WOH\,G457, the MIPS-SED
pointing differs by $23^{\prime\prime}$. This region contains several
molecular clouds, and an association with an optically inconspicuous compact
dust cloud or YSO is perhaps more likely. The same may be true for the nearby,
anonymous MIPS-SED target UFO\,1.

\subsection{MSX-LMC\,1794 (\#46)}

At the edge of a more extended H\,{\sc ii} region, this is an unremarkable
(ultra)compact H\,{\sc ii} source except for the fact (which we note here)
that in the {\it Spitzer} IRS spectrum (Buchanan et al.\ 2006) the 17-$\mu$m
PAH complex is relatively strong and the continuum emission in the 20--35
$\mu$m range resembles a power-law more than a dust emission continuum.

\subsection{MSX-LMC\,956 (\#47)}

The IR source appears as a bright knot on the rim of the H\,{\sc ii} region
N\,176 (Henize 1956) = DEM\,L\,280 (Davies et al.\ 1976) (cf.\ Indebetouw et
al.\ 2008). Somewhat further West lie two molecular clouds, viz.\
30\,Dor\,Center\,6 (Kutner et al.\ 1997) and 30\,Dor-C07 (Caldwell \& Kutner).
Though both named central to 30\,Dor, they are really part of the {\it
Southern} molecular ridge distinguished from the 30\,Dor giant H\,{\sc ii}
region (the Tarantula Nebula).

\subsection{BSDL\,2959 (\#48)}

Maybe associated with IRAS\,05458$-$6710, BSDL\,2959 is one of a pair (with
BSDL\,2956) of small star clusters with associated nebulosity (Bica et al.\
1999), located within the N\,74A H\,{\sc ii} region (Henize 1956).

\section{Results}

\begin{figure*}
\epsscale{1.17}
\vbox{
\plotone{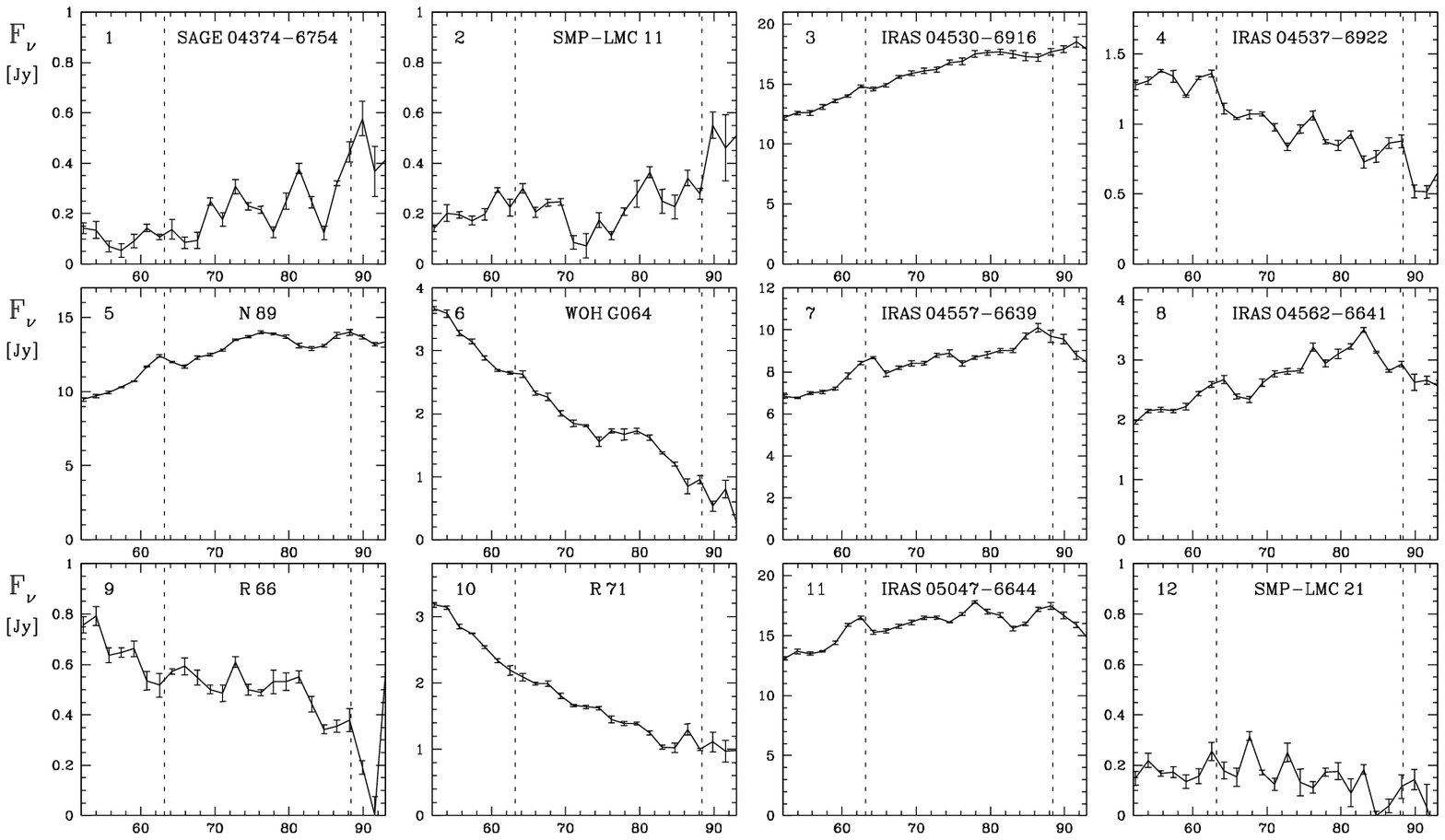}
\plotone{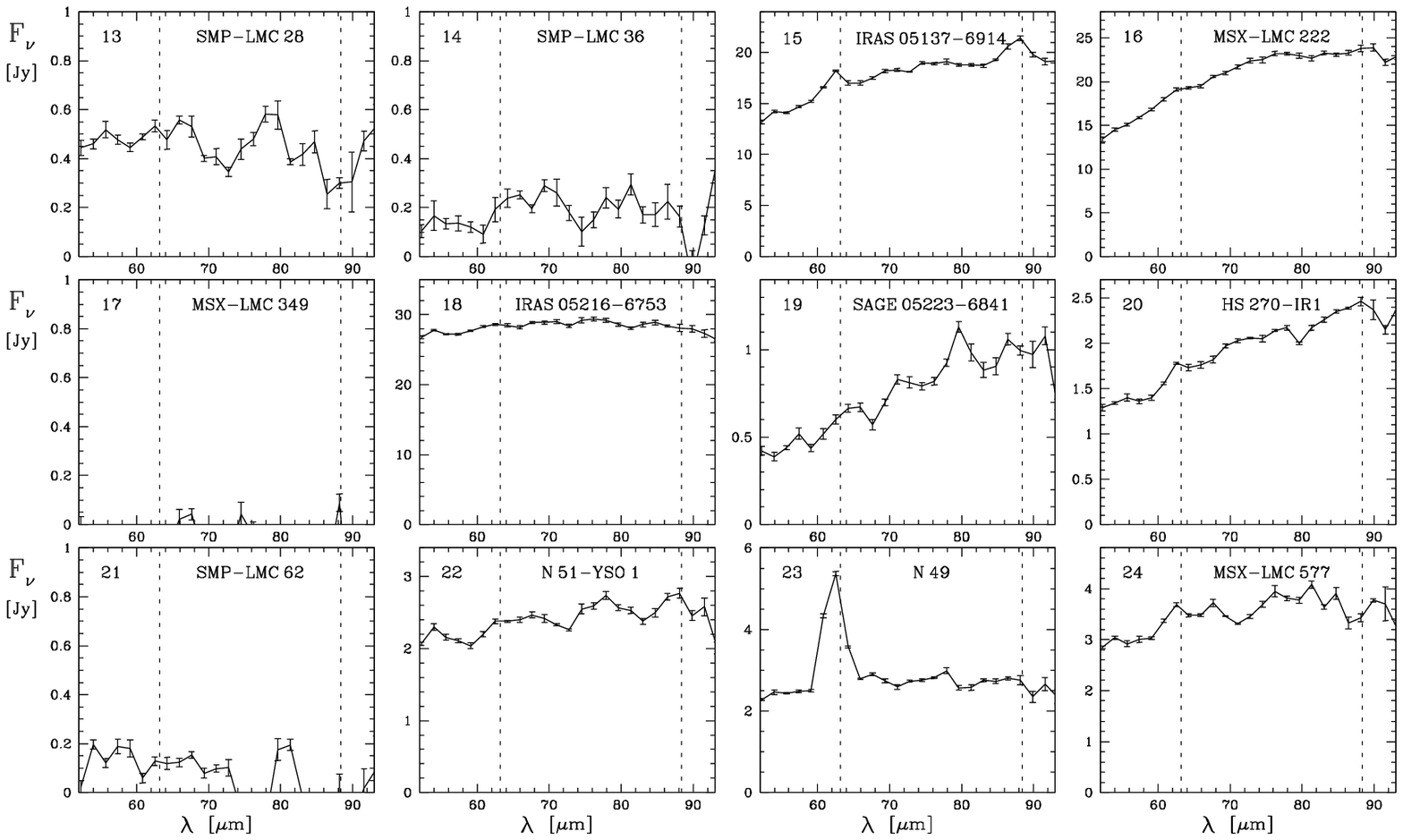}
}
\caption{MIPS-SED spectra of all 48 targets. Vertical dashed lines indicate
the positions of the [O\,{\sc i}] and [O\,{\sc iii}] fine-structure lines at
$\lambda=63$ and 88 $\mu$m, respectively.}
\label{f4}
\end{figure*}

\setcounter{figure}{3}

\begin{figure*}
\epsscale{1.17}
\vbox{
\plotone{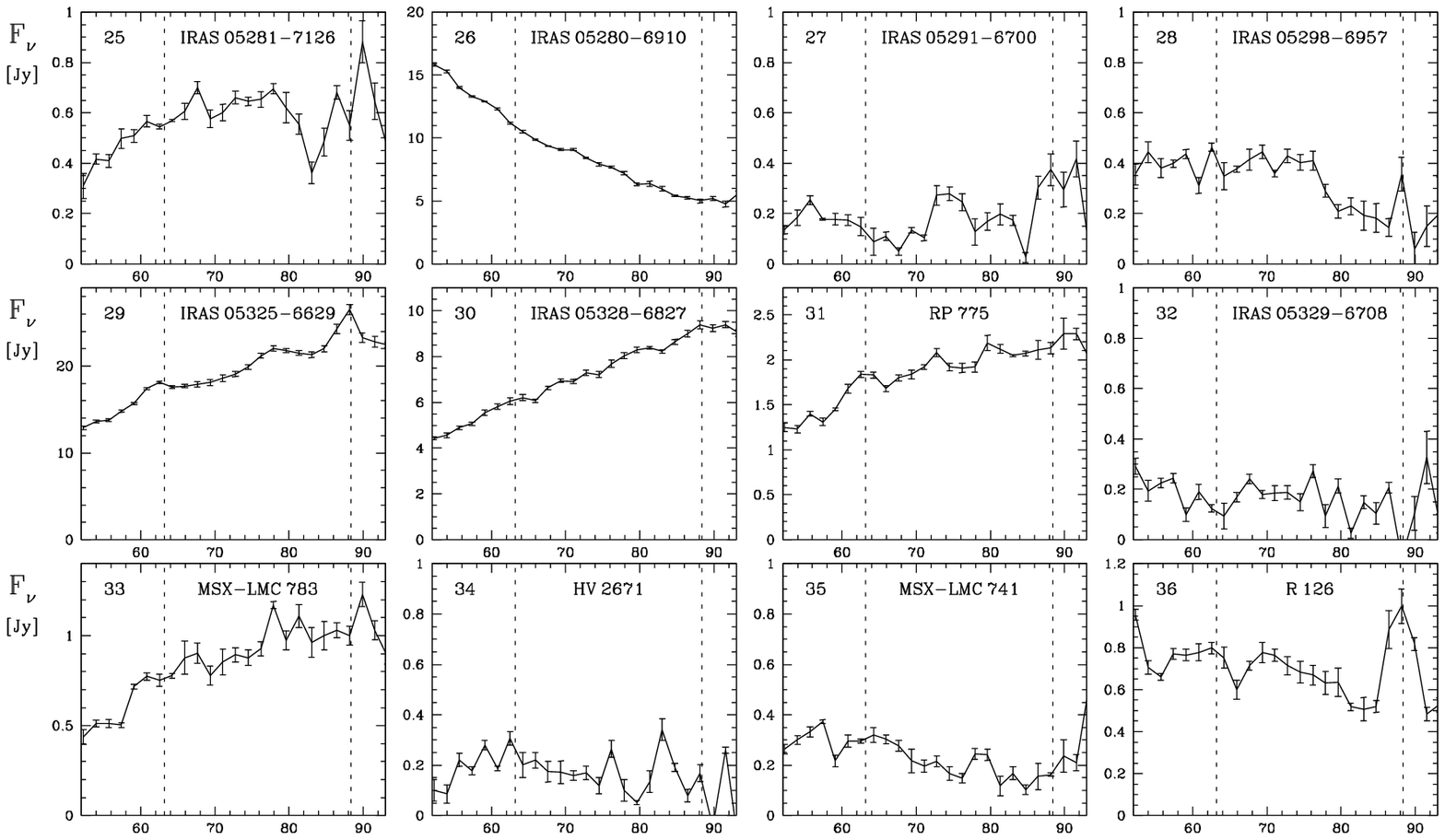}
\plotone{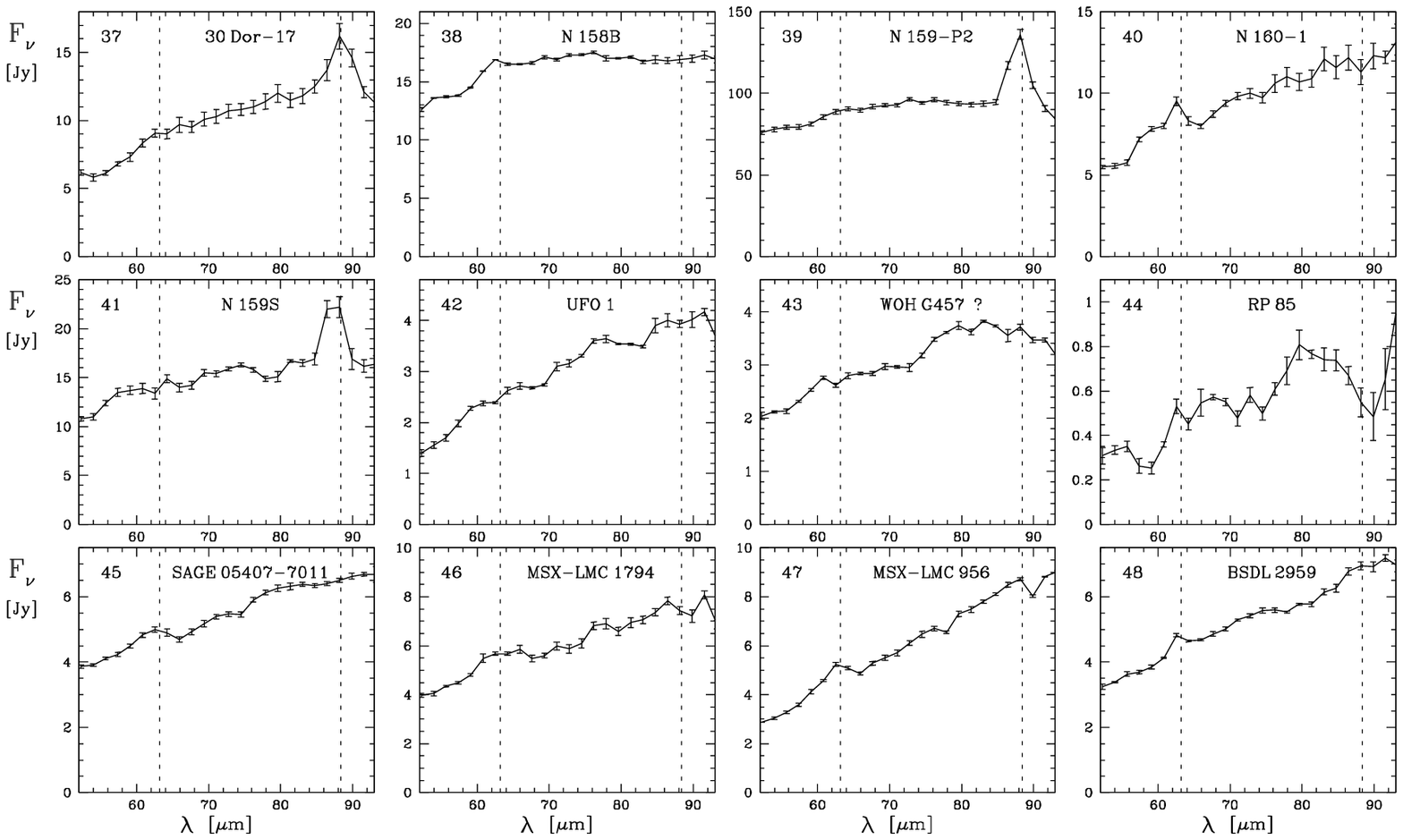}
}
\caption{Continued.}
\label{f4}
\end{figure*}

The MIPS-SED spectra of all 48 targets are presented in Figure 4. Four sources
have poor or bad spectra. One of these, MSX-LMC\,349, was not detected (a
spectrum was extracted nonetheless, but no trace of the target could be
detected in the 2-D frames); it is clearly detected in the broad MIPS
70-$\mu$m band, but at a level similar to the structured background (see Fig.\
2). Table 4 summarizes properties derived from the {\it Spitzer} data.

Often, one or two fine-structure emission lines can be seen, [O\,{\sc i}]
$^3$P(1--2) and [O\,{\sc iii}] $^3$P(0--1), at $\lambda=63.2$ and 88.4 $\mu$m,
respectively. These are discussed in \S 5.1. There is no convincing detection
of the [N\,{\sc iii}] $^2$P(1/2--3/2) transition at $\lambda=57.3$ $\mu$m, and
we discuss the implication in \S 5.2. There is evidence for additional
discrete features in the spectra of some objects, but their identification is
uncertain. They are discussed in \S 5.4. The slope of the continuum is
generally either steeply rising (most common) or steeply declining (less
common); this is an indication of differences in dust temperature and is
discussed in \S 5.3.

\subsection{Clarification of the nature of selected objects}

\subsubsection{IRAS\,04537$-$6922/GV\,60 (\#4), WOH\,G457 (\#43)}

The MIPS-SED spectra of IRAS\,04537$-$6922 and WOH\,G457 are definitely not
typical of M-type giants with warm circumstellar dust, and therefore unlikely
to be due to GV\,60 and WOH\,G457, respectively. We confirm that the MIPS-SED
spectrum of IRAS\,04537$-$6922 is that of the WR-type star Brey\,3a. The true
MIPS-SED source associated with the WOH\,G457 pointing is likely a YSO (see
Indebetouw et al.\ 2008).

\subsubsection{The B[e] star IRAS\,04530$-$6916 (\#3)}

The question here is, whether this B[e] star is an evolved object or a young,
embedded star. The MIPS-SED spectrum looks very much like that of YSOs, with a
steeply rising dust continuum and an [O\,{\sc i}] emission line. This is very
different from the MIPS-SED spectra of the known evolved B[e] stars in our
sample, R\,66 and R\,126, that have a declining dust continuum. We thus
conclude that IRAS\,04530$-$6916 is young, not an evolved star.

\subsubsection{The very low excitation nebula N\,89 (\#5)}

This object has a MIPS-SED spectrum very much like that of the young B[e] star
IRAS\,04530$-$6916 described above, and we thus tentatively conclude that
N\,89 too is a young object.

\subsubsection{The (proto-)Planetary Nebulae (candidates): SMP-LMC\,11 (\#2),
-21 (\#12), -28 (\#13), -36 (\#14), and -62 (\#21), and RP\,85 (\#44) and
IRAS\,05047$-$6644 (\#11)}

These objects show very similar and characteristic MIPS-SED spectra, with a
rather faint and flat dust continuum and no --- or weak --- emission lines.
This includes SMP-LMC\,62, which lacks evidence for either strong shocks
(strong [O\,{\sc i}], see \S 5.1.1) or a high electron density (strong
[O\,{\sc iii}]) and which might therefore not be so extra-ordinary but a
normal PN. It also includes RP\,85, confirming its likelihood of being a
normal PN.

However, the exception is IRAS\,05047$-$6644, whose PN nature was in doubt for
its huge size and super-AGB luminosity. The MIPS-SED spectrum of this source
is totally different from the PNe, with a rising continuum and strong emission
lines. We suggest this object is more likely to be a luminous, i.e.\ massive,
object and not a genuine PN.

\subsubsection{The H\,{\sc ii} blobs: IRAS\,05137$-$6914 (\#15),
IRAS\,05325$-$6629 (\#29), RP\,775 (\#31), N\,158B (\#38), and MSX-LMC\,1794
(\#46), and the nebulous star cluster BSDL\,2959 (\#48)}

These (ultra?)compact H\,{\sc ii} regions have similar spectra, with a rising
dust continuum and one or two emission lines. They are probably young, massive
stars in the process of sculpting an ionized cavity inside their dust cocoons.

The dust in N\,158B seems somewhat warmer, and together with the rather strong
[O\,{\sc i}] line its MIPS-SED spectrum resembles more that of
IRAS\,05047$-$6644. Hence, we classify N\,158B as a high-mass star, but are
undecided about its youthfulness.

\subsubsection{The unclassified sources: MSX-LMC\,577 (\#24),
IRAS\,05281$-$7126 (\#25), MSX-LMC\,741 (\#35), UFO\,1 (\#42),
SAGE\,05407$-$7011 (\#45), and MSX-LMC\,956 (\#47)}

These six sources lack classification in the literature. The latter three are
almost certainly YSOs (see Indebetouw et al.\ 2008). The MIPS-SED spectra of
the first two also resemble YSOs.

The spectrum of MSX-LMC\,741 is rather noisy as the object is faint at 70
$\mu$m. The MIPS photometry and MIPS-SED spectral slope both suggest a lack of
cold dust. It is in fact similar to the high-mass stellar object
IRAS\,05216$-$6753; we concur that both are likely to represent mature or
evolved evolutionary stages.

\subsubsection{The extra-galactic nature of SAGE\,04374$-$6754 (\#1) and
SAGE\,05223$-$6841 (\#19)}

Two of the three SAGE-named objects show a rising continuum but no conspicuous
emission lines at the expected positions (hence their classification as C0,
see \S 4.3). At least one of them, SAGE\,04374$-$6754 is a background galaxy:
the {\it Spitzer} IRS spectrum of this isolated, very red object, which is
seen toward the outskirts of the LMC, undoubtedly reveals a redshift of
$z\approx0.175$ (Woods et al., in prep.). We speculate that the other,
SAGE\,05223$-$6841 is a background galaxy too. In most galaxies the [O\,{\sc
i}] line at 63 $\mu$m is brighter than the [O\,{\sc iii}] line at 52 $\mu$m,
e.g., in the nearby starburst galaxy M\,82 (Colbert et al.\ 1999), the giant
elliptical Active Galactic Nucleus Cen\,A (Unger et al.\ 2000), and the normal
spiral galaxy NGC\,4414 (Braine \& Hughes 1999). The ratio of the [O\,{\sc i}]
line flux to the far-IR flux is typically just over $10^{-3}$, while for the
[O\,{\sc iii}] line (at 52 $\mu$m) it is usually a few times less than that
(Negishi et al.\ 2001), though not always (cf.\ Lord et al.\ 1996). If the
line at 80 $\mu$m observed in the MIPS-SED spectrum of SAGE\,05223$-$6841 is
the redshifted [O\,{\sc i}] line then the redshift would be $z\approx0.27$; if
it is the [O\,{\sc iii}] line then it would be $z\approx0.54$.

\subsection{Serendipitous spectra}

\begin{figure*}
\epsscale{1.17}
\plotone{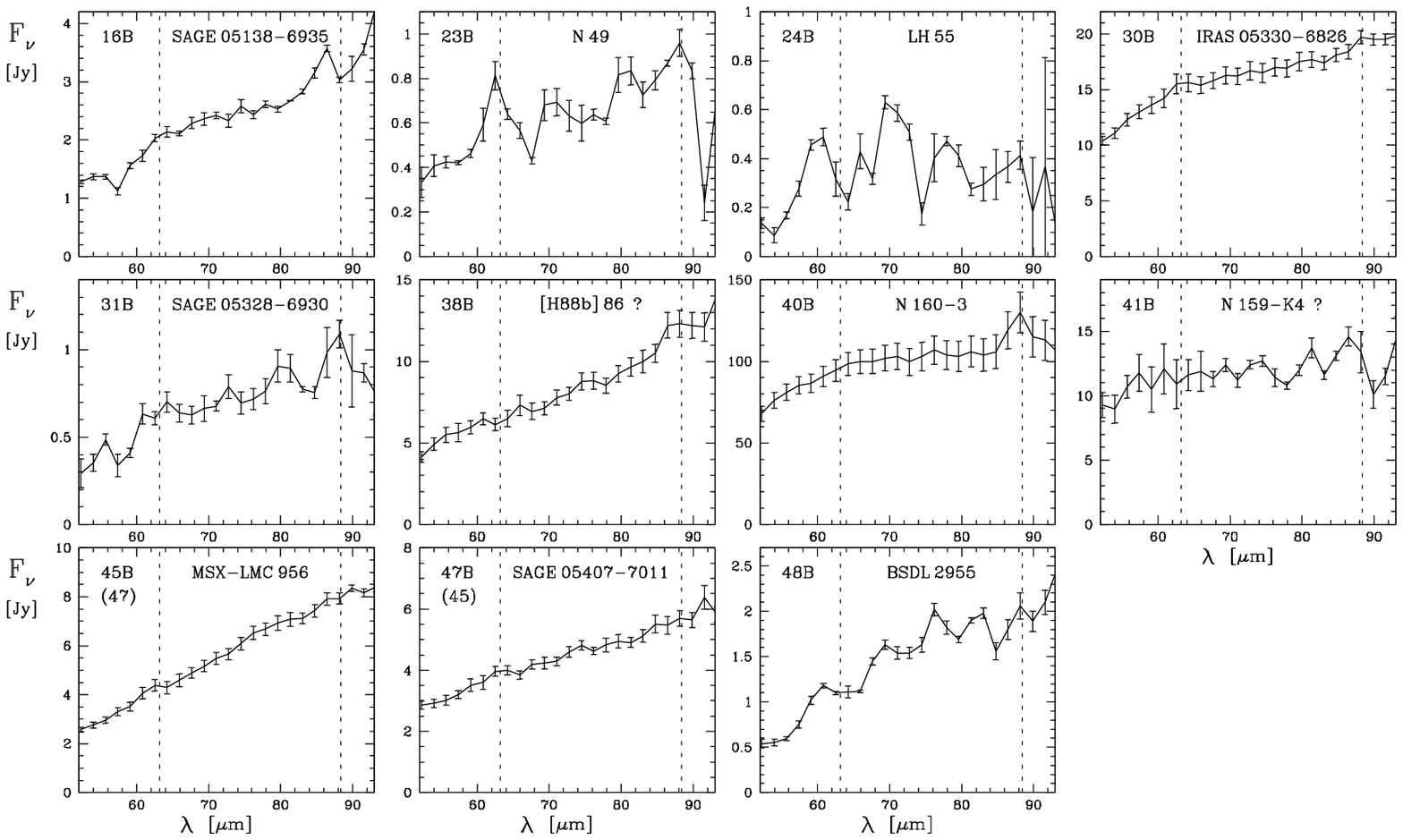}
\caption{Serendipitous MIPS-SED spectra. Vertical dashed lines indicate the
positions of the [O\,{\sc i}] and [O\,{\sc iii}] fine-structure lines at
$\lambda=63$ and 88 $\mu$m, respectively.}
\label{f5}
\end{figure*}

A number of additional spectra could be extracted from the 2-D frames (Table 5
and Fig.\ 5). These sources were not generally well-centered within the slit,
which will have led to flux losses --- no attempt was made to correct for
this. In three instances the serendipitous spectra were of principal targets,
viz.\ N\,49 (although a different portion of the SNR), SAGE\,05407$-$7011, and
MSX-LMC\,956. The spectrum of this part of N\,49 does not exhibit as strong an
[O\,{\sc i}] line as the principal target position. The spectra of the other
two re-observed sources are consistent with the original observations.

\subsubsection{LH\,55 (\#24B)}

The stellar association LH\,55 (Lucke \& Hodge 1970) sits in an H\,{\sc ii}
region from which were detected soft X-rays (Wang \& Helfand 1991) and radio
continuum emission (Filipovi\'c, Jones \& White 2003, and references therein).

\subsubsection{IRAS\,05330$-$6826 (\#30B)}

This IR source (also MSX-LMC\,770) is likely a compact H\,{\sc ii} region
(Egan et al.\ 2001; Kastner et al.\ 2008) and a weak source of radio continuum
emission (\#49 in Marx, Dickey \& Mebold 1997). It was imaged in the peak-up
array while acquiring the IRS spectrum of IRAS\,05328$-$6827 (van Loon et al.\
2005c). The MIPS-SED spectrum is of good quality, and is characterized by a
cold continuum on top of which there is some line emission from [O\,{\sc i}]
and [O\,{\sc iii}] (Fig.\ 5).

\subsubsection{[H88b]\,86 (\#38B)}

This IR source is possibly associated with the dark cloud [H88b]\,86 (Hodge
1988), measuring $26\times12$ pc$^2$. This is commensurate with the MIPS-SED
spectrum which is rather featureless and steeply increasing (Fig.\ 5),
suggesting a quiescent, cold dust cloud.

\subsubsection{N\,160-3 (\#40B)}

The MIPS-SED spectrum (Fig.\ 5) of the compact far-IR source within this
molecular cloud resembles that of (ultra)compact H\,{\sc ii} regions, with a
clear [O\,{\sc iii}] line.

\subsubsection{N\,159-K4 (\#41B)}

This near-IR source is only moderately red (Gatley, Hyland \& Jones 1982); the
MIPS-SED spectrum (Fig.\ 5) is rather flat, suggesting that N\,159-K4 may be a
hot, massive star inside a molecular cloud.

\subsubsection{BSDL\,2955 (\#48B)}

The association BSDL\,2955 (also MSX-LMC\,1432) is in fact closer to
IRAS\,05458$-$6710 (and brighter at 70 $\mu$m) than the MIPS-SED principal
target, BSDL\,2959.

\subsection{A simple MIPS-SED classification scheme}

The limits in resolution and spectral range result in few conspicuous spectral
features that nonetheless vary considerably between sources. It thus remains
meaningful to devise a simple classification scheme, based on the spectral
appearance. We dub this ``The Keele System''. The primary spectral type is
defined as follows:
\begin{itemize}
\item[$\bullet$]{An upper-case letter denotes the continuum slope, for a
spectrum expressed in F$_\nu$ as a function of $\lambda$: C = rising (e.g.,
cold dust); F = flat (this includes spectra with a peak mid-way the MIPS-SED
range); W = declining (e.g., warm dust);}
\item[$\bullet$]{Following the upper-case letter, a number denotes the
presence of the [O\,{\sc i}] and [O\,{\sc iii}] lines: 0 = no oxygen lines are
present; 1 = the [O\,{\sc i}] line is present, but the [O\,{\sc iii}] line is
not; 2 = both [O\,{\sc i}] and [O\,{\sc iii}] lines are present; 3 = the
[O\,{\sc iii}] line is present, but the [O\,{\sc i}] line is not.}
\end{itemize}
A secondary classification is based on additional features: a lower-case
letter ``b'' may follow the primary type in the presence of a bump in the
$\lambda\sim70$--80 $\mu$m region.

Clearly, the primary type can be a diagnostic of the dust temperature and the
excitation conditions in the gas. The secondary type has less immediate
diagnostic value, but it may be used to isolate special classes of objects.
This can be explored by investigating how the spectral types vary among the
different classes of targets, and the degree at which secondary and primary
subclasses are correlated. This we shall do in \S 5.

We have classified all MIPS-SED spectra (Tables 4 \& 5). We have erred on the
side of caution with respect to the detection of spectral lines. So, for
instance, an object with spectral type C0 may still display weak oxygen lines
in a higher-quality spectrum..

\section{Discussion}

We first discuss the prominent oxygen fine-structure emission lines (\S 5.1),
followed by the nitrogen line (\S 5.2), and then the dust continuum (\S 5.3)
and discrete features possibly due to ice, molecules or minerals (\S 5.4). At
the end of this section, we summarize the population MIPS-SED characteristics
(\S 5.5).

\subsection{Oxygen}

The [O\,{\sc i}] and [O\,{\sc iii}] lines at $\lambda=63$ and 88 $\mu$m,
respectively, are of great diagnostic value with regard to the excitation
conditions of the gas within the objects, even if these lines are absent.
Unfortunately, the [O\,{\sc iii}] $^3$P(1--2) transition at $\lambda=51.8$
$\mu$m is just outside the spectral range covered by MIPS-SED; the [O\,{\sc
i}]$_{63}$/[O\,{\sc iii}]$_{52}$ flux ratio is a good measure of the electron
density (e.g., Liu et al.\ 2001), but the [O\,{\sc i}]$_{63}$/[O\,{\sc
iii}]$_{88}$ line ratio can also be used for that (Giannini et al.\ 2000;
Rubin et al.\ 1994). The line luminosities can be found in Table 4, for an
assumed distance of 50 kpc. They were computed by summing the three spectral
points centered on the line after subtracting a continuum obtained by linear
interpolation between the spectral points immediately adjacent to the
integration interval. The error was computed by adding in quadrature the
errors in the three spectral points that were summed, and three times the
error in the mean of the two continuum anchors (to account for the error in
the continuum estimate at each of the three spectral points).

We first describe the use of the [O\,{\sc i}] line as a diagnostic of fast
shocks, in the context of its prominence in the MIPS-SED spectrum of the only
known SNR in our sample, N\,49. Then, we discuss the appearance of the lines,
first in H\,{\sc ii} regions, molecular clouds, and YSOs within them, and then
in evolved objects. After that, we assess their contribution to broadband
photometry, which will be useful in interpreting the MIPS photometric
properties of objects for which no spectral information is available.

\subsubsection{Shocks versus photon-dominated regions}

The [O\,{\sc i}] line is the main cooling transition in the dense post-shocked
regions of dissociative J-type shocks (e.g., Giannini, Nisini \& Lorenzetti
2001; cf.\ Hollenbach \& McKee 1989). Whereas in C-type shocks it is only a
minor coolant, H$_2$ being the dominant coolant (Le Bourlot et al.\ 2002).
J-type shocks are more powerful ($v\sim100$ km s$^{-1}$) and associated with
strong sources of feedback, C-type shocks ($v\approx30$ km s$^{-1}$) are more
typical of the ambient ISM. So, in a SNR for instance a J-type shock might be
anticipated. Indeed, this was observed in the SNR IC\,443, which is seen to be
interacting with a molecular cloud (Burton et al.\ 1990; Rho et al.\ 2001). In
the absence of such strong interaction the [O\,{\sc i}] line is fainter than
the [O\,{\sc iii}] line, as observed in the Crab Nebula (Green, Tuffs \&
Popescu 2004) and in SN\,1987A (Lundqvist et al.\ 1999).

With an excitation temperature of $\Delta E/k=228$ K, the [O\,{\sc i}] line
also forms in photon-dominated regions (PDRs), such as the interface between
an H\,{\sc ii} region and a molecular cloud (Giannini et al.\ 2000; cf.\
Tielens \& Hollenbach 1985). It is the main coolant of infalling envelopes
around YSOs (Ceccarelli, Hollenbach \& Tielens 1996). In contrast, the
[O\,{\sc iii}] line is strong in highly-ionized diffuse gas (Mizutani, Onaka
\& Shibai 2002), as observed with the KAO in 30\,Dor by Lester et al.\ (1987).

\subsubsection{Shocks in the SNR N\,49}

The only known SNR in our sample, N\,49 exhibits a phenomenally strong
[O\,{\sc i}] line, by far the strongest with respect to the continuum in our
entire sample. This is an impressive confirmation of the work of strong shocks
in the interaction interface of the SNR with an adjacent molecular cloud. It
is testimony of the destructive effect SNe have on dust in the surrounding
ISM. One could anticipate N\,49 to be a bright source of $\gamma$-rays,
resulting from particle acceleration in these shocks.

\subsubsection{Oxygen in star-forming regions and YSOs}

Ultra-compact H\,{\sc ii} regions always show the [O\,{\sc i}] line, but not
always the [O\,{\sc iii}] line (Peeters et al.\ 2002). And though the former
generally outshines the latter, there are cases in which the [O\,{\sc iii}]
line is dominant. For instance, Lerate et al.\ (2006) found in the Orion\,KL
region [O\,{\sc i}] and [O\,{\sc iii}] to be comparatively similar in
strength; Giannini et al.\ (2000) found that in the molecular cloud associated
with NGC\,2024, [O\,{\sc i}] is more than ten times brighter than [O\,{\sc
iii}]. Higdon et al.\ (2003) found that H\,{\sc ii} regions in M\,33 exhibit
[O\,{\sc iii}] lines of similar strength to [O\,{\sc i}] or (much) stronger,
e.g., in NGC\,604. Thus, there may exist a chronological sequence from
[O\,{\sc i}]-dominated molecular clouds to [O\,{\sc iii}]-dominated giant
H\,{\sc ii} regions.

[O\,{\sc i}] is observed in dark clouds, due to shocks (Nisini et al.\ 1999a).
Although some embedded outflow sources exhibit a rather featureless spectrum
(Froebrich et al.\ 2003), weak [O\,{\sc i}] emission is seen on top of the
cold continuum of Class O protostar L\,1448-mm (Nisini et al.\ 1999b;
Ceccarelli et al.\ 1998). Van den Ancker, Tielens \& Wesselius (2000) compared
the flat-continuum sources S\,106\,IR and Cep\,A\,East: [O\,{\sc i}] is strong
in S\,106\,IR but very weak in Cep\,A\,East; the former is dominated by a PDR
(cf.\ Schneider et al.\ 2003) whereas the latter is heavily embedded.

Shocked gas was found in the vicinity of the pre-main sequence system T\,Tauri
(Van den Ancker et al.\ 1999). Van den Ancker, Wesselius \& Tielens (2000)
showed that in Herbig Ae/Be stars, [O\,{\sc i}] is considerably stronger than
[O\,{\sc iii}] (cf.\ Lorenzetti et al.\ 2002). However, [O\,{\sc iii}] is
sometimes seen in Herbig Ae/Be stars (e.g., V645\,Cyg; Lorenzetti et al.\
1999), and also in the outflow protostar TC\,4 in the Trifid Nebula (Lefloch
\& Cernicharo 2000); it is stronger than [O\,{\sc i}] also in the molecular
cloud core M\,17-North (Henning et al.\ 1998). This may be due to massive
protostars already carving out an ultra-compact H\,{\sc ii} region whilst
still heavily embedded. We thus take the [O\,{\sc iii}] line as a diagnostic
of (ultra-)compact H\,{\sc ii} regions surrounding massive stars, and the
[O\,{\sc i}] line as a diagnostic of shocks or PDRs in earlier stages or less
massive protostars.

The YSOs 30\,Dor-17 and N\,159-P2 are exquisite examples of intense [O\,{\sc
iii}] emitters. These must contain compact H\,{\sc ii} regions, with shocks
due to molecular outflows only playing a minor r\^ole.

\subsubsection{Oxygen in evolved objects}

Haas \& Glassgold (1993) and Haas, Glassgold \& Tielens (1995) detected
[O\,{\sc i}] in the famous RSGs Betelgeuse, Antares and Rasalgethi
($\alpha$\,Herculis), and they argued that the line is formed in the inner
part of the dense wind, roughly where dust condenses. The line would therefore
be a very useful probe of this critical region in the outflow. In our sample,
the line is indeed visible in the cool, very luminous RSG WOH\,G064 (see \S
5.1.5), but not in any of the other OH/IR stars in our sample.

In objects evolving beyond a cool, dusty phase, such as proto-planetary
nebulae (the early transition stage between AGB and PN), the relative
intensities of the [O\,{\sc i}] and [O\,{\sc iii}] lines may act as a
chronometer: Castro-Carrizo et al.\ (2001) and Fong et al.\ (2001) found that
fine-structure lines are only seen in evolved stars with $T_\star>10,000$ K,
i.e.\ they arise from PDRs not shocked regions. [O\,{\sc i}] is stronger than
[O\,{\sc iii}] in PNe with plenty of (warm) dust, but much the opposite in PNe
lacking (warm) dust (e.g., Bernard-Salas \& Tielens 2005). For example,
[O\,{\sc i}] is seen on top of a warm continuum in the proto-planetary nebula
IRAS\,16594$-$4656 (Garc\'{\i}a-Lario et al.\ 1999), and even in the
carbon-rich proto-planetary nebula CRL\,618 (Herpin \& Cernicharo 2000); on
the other hand, no fine-structure lines were detected in the carbon-rich
proto-planetary nebula AFGL\,2688 (the Egg Nebula; Cox et al.\ 1996) which has
a warm continuum too.

Interestingly, of the seven (proto-)PNe in our sample, four are classified F0,
whilst the spectra of the others are slowly-rising but no or hardly any trace
of oxygen line emission; the only exception being IRAS\,05047$-$6644 which
shows cold dust and both oxygen lines (and an additional feature at 78
$\mu$m). The latter may not be a PN, though (see \S 3.10). No oxygen lines are
detected in any of the four carbon-rich objects in our sample.

\subsubsection{Detection of oxygen fine-structure line emission in the RSG
WOH\,G064}

With a line flux of $F({\rm [OI]})\approx2\times10^{-16}$ W m$^{-2}$, the
[O\,{\sc i}] line in WOH\,G064 is only $\approx100$ times as dim as the line
in Betelgeuse and Antares, which are only $\approx200$ pc away rendering the
line in WOH\,G064 $\approx600$ times as powerful. This may be due in part to
the much higher mass-loss rate of WOH\,G064, $\dot{M}\sim10^{-3}$ M$_\odot$
yr$^{-1}$ (van Loon et al.\ 1999b; Ohnaka et al.\ 2008) compared to a few
$\times10^{-6}$ M$_\odot$ yr$^{-1}$ for Betelgeuse and $\sim10^{-5}$ M$_\odot$
yr$^{-1}$ for Antares (Haas et al.\ 1995). But this still leaves unexplained a
factor $\sim3$--6 difference. The dust-free zones in metal-poor envelopes
appear to extend further than those in metal-richer, dustier envelopes (van
Loon et al.\ 2005a), and the outflow velocities are lower leading to higher
densities in the wind (Marshall et al.\ 2004); this would help explain the
bright line in WOH\,G064.

We may speculate that a larger (weakly) ionized region might facilitate
Alfv\'en waves to couple to the circumstellar gas reservoir and thus (help)
drive a wind, where in neutral media Alfv\'en waves would not couple and
quickly damp (see Hartmann \& MacGregor 1980).

Alternatively, we may have sampled emission from the ISM surrounding
WOH\,G064, as in the case of the optical fine-structure lines.

\subsubsection{The contribution of oxygen fine-structure line emission to MIPS
70-$\mu$m broadband photometry}

The MIPS 70-$\mu$m band peaks around $\lambda=71$ $\mu$m, and the filter
response function, $S_\lambda$, drops to half the peak value at $\lambda=61$
and $\lambda=80$ $\mu$m. At the positions of the [O\,{\sc i}] and [O\,{\sc
iii}] lines, $S_{63.2}\approx0.73\ S_{71}$ and $S_{88.4}\approx0.145\ S_{71}$,
respectively. Thus, in particular a very strong [O\,{\sc i}] line can make a
considerable contribution to the total flux in the MIPS 70-$\mu$m band. To
assess the extent to which this is the case, we determined the ratio of the
bandpass-convolved [O\,{\sc i}] line flux to the bandpass-convolved integrated
MIPS-SED spectrum.

The SNR N\,49 has a strong contribution from [O\,{\sc i}]: 11\% of the MIPS
70-$\mu$m flux is due to this line. We thus caution the use of the 70-$\mu$m
band to determine dust content of SNRs --- as is the case for the 24-$\mu$m
band. Still, the effect is modest even for the intense line in N\,49. The next
most extreme [O\,{\sc i}] emitters (in terms of line-to-continuum ratio),
IRAS\,05047$-$6644, IRAS\,05137$-$6914 and N\,158B are only affected at the
1\% level. The most extreme [O\,{\sc iii}] emitters, N\,159-P2 and 30\,Dor-17
are only affected by the [O\,{\sc iii}] line emission by 1\%, in spite of
their huge line-to-continuum ratio of about two.


\subsection{Nitrogen}

The ionization potentials of the first four stages of nitrogen and oxygen are
very similar --- 14.5/13.6, 29.6/35.1, 47.4/54.9, and 77.5/77.4 eV
for N$^0$/O$^0$, N$^+$/O$^+$, N$^{2+}$/O$^{2+}$, and N$^{3+}$/O$^{3+}$,
respectively --- and thus the nitrogen-to-oxygen abundance ratio $N({\rm
N})/N({\rm O})\approx N({\rm N}^{2+})/N({\rm
O}^{2+})=(F_{57}/j_{57})/(F_{88}/j_{88})$ (Lester et al.\ 1987; Simpson et
al.\ 1995; Liu et al.\ 2001). The emissivity ratio increases from
$j_{57}/j_{88}\approx1.4$ at $n_{\rm e}\la10^2$ cm$^{-3}$ to $\approx6$ at
$n_{\rm e}\ga10^5$ cm$^{-3}$ (Liu et al.\ 2001).

PNe range from $N({\rm N})/N({\rm O})\approx0.1$ to $\approx2$, where 0.12 is
the solar value (Dinerstein et al.\ 1995; Rubin et al.\ 1997; Liu et al.\
2001). Lester et al. (1987) used KAO data to perform measurements also in
30\,Dor; $N({\rm N})/N({\rm O})=0.14$--0.40 in Galactic H\,{\sc ii} regions
but $\approx0.06$ in 30\,Dor (Rubin et al.\ 1988; Simpson et al.\ 1995).
Rudolph et al.\ (2006) summarize similar results: in ultra-compact H\,{\sc ii}
regions in the central regions of the Milky Way, $N({\rm N}^{2+})/N({\rm
O}^{2+})\ga0.5$, but it decreases to $<0.2$ beyond Galactocentric distances of
$\approx14$ kpc (Rudolph et al.\ 1997); values in the LMC are $\approx0.04$
and even lower in the SMC (Roelfsema et al.\ 1998).

Can we place interesting upper limits? [O\,{\sc iii}] is faint in most of our
targets, so it would be difficult to detect the [N\,{\sc iii}] line. But in
the objects with the strongest [O\,{\sc iii}] line, 30\,Dor-17 and N\,159-P2,
we would have expected to detect the [N\,{\sc iiii}] line at $>0.1$ times the
continuum. This is clearly not the case. Recognizing that a 3-$\sigma$
detection would amount to $\approx1/6^{\rm th}$ of the flux in the [O\,{\sc
iii}] line in N\,159-P2, we place a firm upper limit of $N({\rm N})/N({\rm
O})\la0.1$, but quite possibly as low as $<0.03$ as the strong [O\,{\sc iii}]
line compared to the dust continuum indicates a high electron density.

\subsection{Dust continuum}

We define the continuum slope as follows:
\begin{equation}
\alpha \equiv 2.44\ \frac{F_\nu(85)-F_\nu(55)}{F_\nu(85)+F_\nu(55)},
\end{equation}
such that $\alpha=0$ in a flat $F_\nu$ spectrum and $\alpha=-1$ in the
Rayleigh-Taylor approximation of a relatively hot Planck curve. These values
were computed from the mean values of the three spectral points centered at 55
and 85-$\mu$m, respectively, and are listed in Tables 4 \& 5. The error was
computed by propagation of the standard deviations in the two sets of three
spectral points used.

The MIPS-SED continua of the objects of our study are due to emission from
relatively cool dust (typically 20--200 K). The 70-$\mu$m band is dominated by
emission from big grains, 15--110 nm in size, in contrast to the 24-$\mu$m
band which is dominated by the emission from very small grains (VSGs; if
present), 1.2--15 nm in size (D\'esert, Boulanger \& Puget 1990; Galliano et
al.\ 2003). The MIPS-SED spectra can not yield reliable dust masses, but
far-IR emission or the lack thereof can place constraints on the production of
dust by evolved stars in the distant past. In YSOs and molecular clouds only a
rather meaningless lower limit to the dust mass can be obtained as
contributions from (even) colder dust are not probed adequately. In the
following, we first estimate dust temperatures, before making attempts to
quantify the amount of dust and the timing of its origin.

\subsubsection{Dust temperatures}

\begin{figure}
\epsscale{1.17}
\plotone{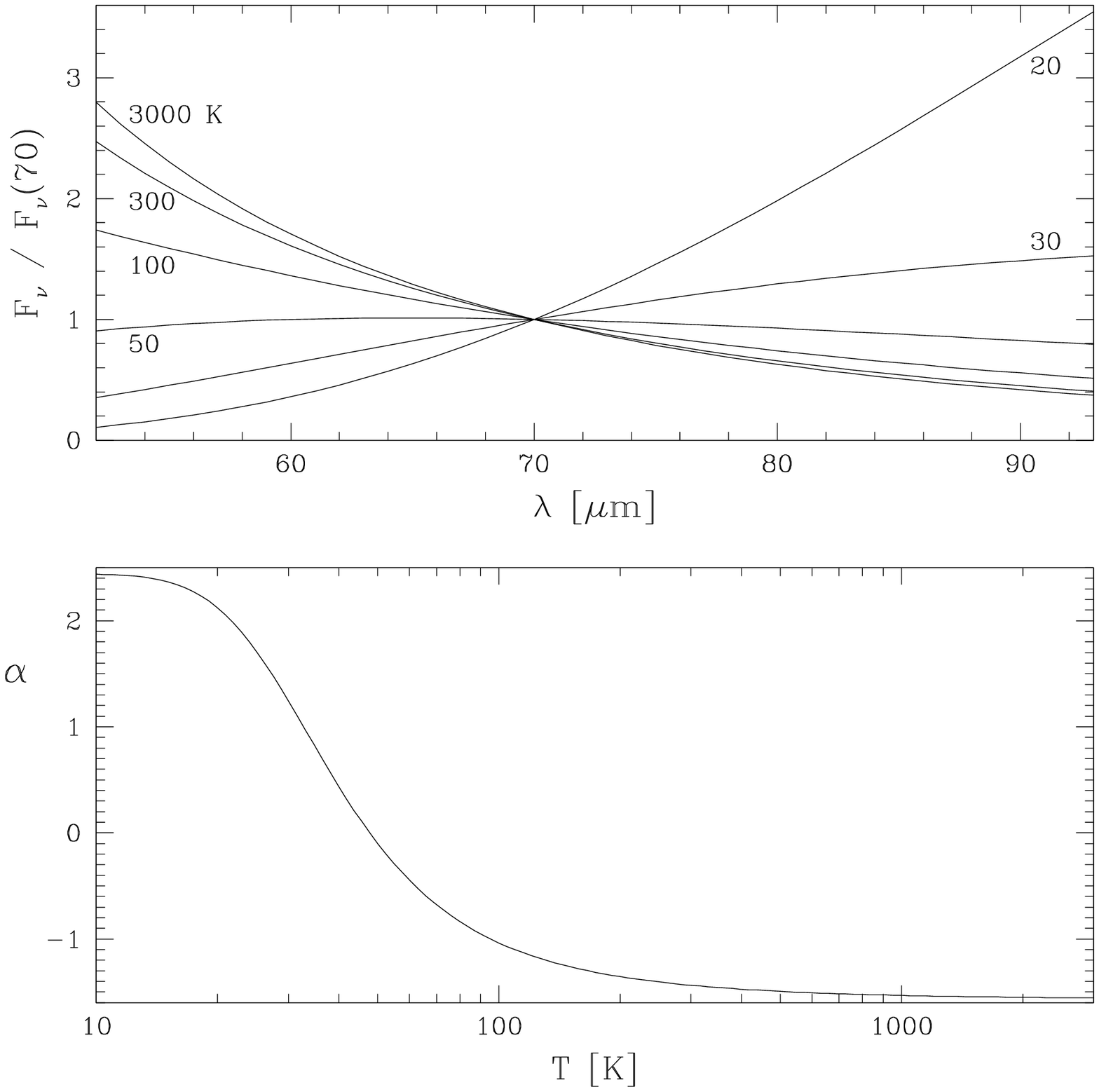}
\caption{{\it Top:} Grey-bodies (Eq.\ 3) for temperatures between 20 and 3000
K. {\it Bottom:} Correlation between grey-body temperature and spectral slope
--- which is defined as: $\alpha \equiv 2.44\
\frac{F_\nu(85)-F_\nu(55)}{F_\nu(85)+F_\nu(55)}$.}
\label{f6}
\end{figure}

The dust temperature characterizing the MIPS-SED range is a powerful
discriminant between warm circumstellar envelopes of evolved stars and cold
molecular clouds in star-forming regions. We have estimated the dust
temperature by comparison with a single grey-body:
\begin{equation}
F_\lambda \propto B_\lambda(T_{\rm dust})\left(1-e^{-\tau_\lambda}\right),
\end{equation}
where $B_\lambda(T_{\rm dust})$ is a Planck emission curve at the dust
temperature $T_{\rm dust}$, the optical depth $\tau_\lambda \propto
\lambda^{-\beta}$, and $\beta=1$--2, here assumed $\beta\equiv1.5$ (cf.\
Goldsmith, Bergin \& Lis 1997) and thus to very good approximation in the
MIPS-SED range $F_\lambda \propto \lambda^{-\beta} B_\lambda$. Representative
curves are plotted in Figure 6, along with a graph of the relation between the
spectral slope $\alpha$ (Eq.\ 2) and temperature. The temperatures estimated
from the spectral slopes are listed in Table 4.

The YSOs and other objects intimately associated with the process of star
formation have temperatures in the range $T_{\rm dust}\approx32$--44 K. This
compares well with H\,{\sc ii} regions in M\,33, which have typically $T_{\rm
dust}=30$--50 K. The temperatures in molecular cloud cores can be lower;
ISO-LWS measurements in the Serpens molecular cloud core yield $T_{\rm
dust}\approx21$--33 K (Larsson et al.\ 2000). In our sample, only the galaxy
SAGE\,04374$-$6754 and the carbon star MSX-LMC\,349 (cf.\ S\ 5.3.2) display
such low (apparent) temperature. Higher temperatures were derived in the
Cygnus\,X cloud DR\,21, $T_{\rm dust}\approx31$--42 K (Jakob et al.\ 2007),
which again is similar to the range in temperatures in our LMC sample of YSOs.

The dust surrounding SN\,1987A is of similar temperature to the YSOs and
compact H\,{\sc ii} regions, viz.\ $T_{\rm dust}\approx37$ K (Lundqvist et
al.\ 1999). The dust in the SNR N\,49 appears to be somewhat warmer, $T_{\rm
dust}\approx45$ K, possibly because the dust is heated in the prominent shocks
that characterize that object.

The RSGs WOH\,G064 and IRAS\,05280$-$6910 have no important contribution from
dust much colder than $T_{\rm dust}<100$ K. This limits the amount of dust
produced in the past (see \S 5.3.2). The same is true for the LBV R\,71.

\subsubsection{The origin of cold dust in evolved objects}

Following Sopka et al.\ (1985), the dust equilibrium temperature in an
optically thin envelope falls off with distance to the photosphere as
$T(r)\propto r^{-\gamma}$, where $\gamma = 2/(4+\beta)$. So, in our case we
assume $T(r)\propto r^{-0.36}$. Applying this to the RSGs WOH\,G064 and
IRAS\,05280$-$6910, assuming typical values for the inner radius of the dust
envelope, $r_0\sim10^{10}$ km, and dust condensation temperature,
$T_0\approx1000$ K, one would estimate dust of a temperature $T=104$--130 K to
be located at a distance $r\sim$ few $\times10^{12}$ km. At a wind speed of
$v\approx 20$ km s$^{-1}$ (Marshall et al.\ 2004) this would have taken a few
thousand years to get there. This strongly suggests that the mass-loss rate
had increased quite dramatically around that time (the onset of the
``superwind''; cf.\ Suh \& Jones 1997, who model the SED resulting from an
enhanced mass-loss episode).

The MIPS-SED data of the other two OH/IR stars, IRAS\,05298$-$6957 and
IRAS\,05329$-$6708, is of lower quality but if the lower dust temperatures of
$T\approx60$--94 K may be believed this could imply a longer duration of the
prolific dust production. Because of their lower luminosity compared to the
RSGs discussed above, the dust condenses closer to these stars in absolute
terms, at $r_0\sim$ few $\times10^8$ km, resulting in a distance
$r\sim10^{12}$ km at which $T\approx60$ K. At a slower wind speed, $v\sim11$
km s$^{-1}$ (Marshall et al.\ 2004), the dust was produced a few thousand
years ago, at a very similar time as the cool dust in WOH\,G064 and
IRAS\,05280$-$6910. This suggests that the duration of the superwind is
several thousand years and does not differ very much between massive AGB stars
and RSGs. Given that the mass-loss rate during this phase scales approximately
with mass (van Loon et al.\ 1999b), this means that in proportion to their
birth mass, RSGs and (massive) AGB stars lose a similar amount of mass during
the superwind phase.

Similar estimates can be made for the LBV R\,71: for a stellar radius of
$\approx100$ R$_\odot$, a dust equilibrium temperature of 107 K is reached at
$r\sim$ few $\times10^{13}$ km. Assuming a constant outflow speed $v\sim100$
km s$^{-1}$, this material was expelled $\sim10^4$ yr ago --- or several times
longer if it originated in a slower RSG wind. This now seems to reconcile the
post-RSG evolutionary timescale with an origin of the cool dust in a RSG
superwind (see \S 3.9) as described above.

The origin of the cold dust apparently associated with the carbon stars,
MSX-LMC\,349, IRAS\,05291$-$6700 and MSX-LMC\,783, is rather puzzling. Its
inference is more based on the 70-$\mu$m photometry than on the MIPS-SED data.
These are indeed faint point sources seen against a relatively bright,
non-uniform background (Fig.\ 2). This suggests that the detected carbon stars
may be seen plowing through relatively dense ISM, and swept-up ISM is the
cause for the far-IR emission rather than dust produced in the winds of these
carbon stars. To some extent this may be true also for the PNe in our sample.

\subsubsection{The amount of swept-up ISM in the SNR N\,49}

To estimate the amount of material associated with the piled-up dust in N\,49,
we first estimate the dust mass following Evans et al.\ (2003), adopting for
the absorption coefficient at $\lambda=70$ $\mu$m a typical value,
$\kappa(70)\approx200$ cm$^2$ g$^{-1}$ (Mennella et al.\ 1998). The
$F_\nu(70)\approx3$ Jy of continuum emission from 45-K dust sampled in N\,49
(Fig.\ 4; 20\% less than the photometric estimate listed in Table 3) thus
corresponds to $M_{\rm dust}\approx0.2$ M$_\odot$. For a gas-to-dust mass
ratio of $\sim500$ this yields $M\sim100$ M$_\odot$ in total --- much less
than that in the adjacent molecular clouds but very similar to the 200
M$_\odot$ of swept-up ISM estimated from X-ray emission (Hughes et al.\ 1998)
and the 150 M$_\odot$ of matter collected from a sphere of 17 pc diameter
filled with 1 particle cm$^{-3}$.

\subsection{Ice, molecules, and minerals}

Roughly half of the sample show an additional discrete spectral feature in the
$\lambda=70$--80 $\mu$m region. This might be due to molecular or solid-state
(dust or ice) material. In this section we assess the possibility that we
detect water ice, molecular emission or minerals.

\subsubsection{Water ice}

Crystalline water ice has a strong band in the MIPS-SED spectral range. The
position varies between $\lambda\approx60$--66 $\mu$m and the feature can
extend beyond 70 $\mu$m (e.g., Malfait et al. 1999; Dijkstra et al.\ 2006).
Apart from YSOs it is also sometimes seen in detached shells around
highly-evolved stars: e.g., the post-AGB objects IRAS\,09371+1212 (Frosty Leo,
Omont et al.\ 1990) and AFGL\,4106 (Molster et al.\ 1999), and the PN
NGC\,6302 (e.g., Barlow 1998) (cf.\ Sylvester et al.\ 1999).

In the low-resolution MIPS-SED data, it can be hard to distinguish between the
water ice band and the[O\,{\sc i}] line at 63 $\mu$m even though the ice band
would be resolved. Unfortunately, we do not have the sharp 44-$\mu$m band for
confirmation, as it falls between the IRS and MIPS-SED coverages. Thus, we can
only reasonably argue for the presence of crystalline water ice if the feature
is strong, so one can distinguish its broad shape from an unresolved atomic
line, or it peaks around $\lambda\approx60$ or 66 $\mu$m, i.e.\ clearly
displaced from the [O\,{\sc i}] line.

This appears to be the case in N\,51-YSO1, MSX-LMC\,577, MSX-LMC\,783, UFO\,1,
and MSX-LMC\,1794. These are all heavily embedded, cold sources associated
with star formation. Interestingly, all these objects also show a broad bump
around $\lambda\approx78$ $\mu$m. CO$_2$ ice was already detected in the {\it
Spitzer} IRS spectrum of N\,51-YSO1 (Seale et al.\ 2009).

The only other YSOs in the MIPS-SED sample that are known to contain ice are
HS\,270-IR1, IRAS\,05328$-$6827 and 30\,Dor-17 (Oliveira et al.\ 2009). There
is no evidence for crystalline water ice in the MIPS-SED spectra of
HS\,270-IR1 or 30\,Dor-17, but the broad hump around 60--66 $\mu$m in
IRAS\,05328$-$6827 might possibly be due to a weak ice feature.

\subsubsection{Molecules}

The MIPS-SED spectral range covers numerous transitions in water (H$_2$O),
hydroxyl (OH) and carbon-monoxide (CO) molecules; see, for instance the line
survey of the Orion\,KL region by Lerate et al.\ (2006). The CO lines are
generally too weak, with stronger lines beyond 100 $\mu$m (Justtanont et al.\
2000). The M-giant R\,Cas shows very weak molecular lines, on top of a warm
dust continuum (Truong-Bach et al.\ 1999). We would not expect to be able to
detect and sufficiently resolve such line emission. However, blends of strong
emission lines could, in principle, appear as features in the MIPS-SED
spectrum of a few-$\mu$m wide.

The strongest gas-phase H$_2$O line often appears at 67 $\mu$m (Harwit et al.\
1998). It is usually seen in conjunction with other lines, e.g., at 79 and 90
$\mu$m (Barlow et al.\ 1996). Strong OH lines are those at 79.15 and 84.51
$\mu$m; the latter is generally the stronger of the two, but not in
IRC\,+10\,420 (Sylvester et al.\ 1997), or Cep\,E (Moro-Mart\'{\i}n et al.\
2001). Again, these lines are usually seen in conjunction with others, such as
those at 65.2 and 71.2 $\mu$m (e.g., Spinoglio et al.\ 2000). Neither these OH
nor H$_2$O lines are generally seen in the presence of [O\,{\sc iii}] (Barlow
et al.\ 1996; Maret et al.\ 2002; Lorenzetti et al.\ 1999); but the o-H$_2$O
3$_{21}$--2$_{12}$ transition at 75.4 $\mu$m (304 K excitation) {\it is} seen
alongside [O\,{\sc iii}], e.g., in bright-rimmed globule IC\,1396\,N (Saraceno
et al.\ 1996) and in Cep\,E (Moro-Mart\'{\i}n et al.\ 2001).

It might be possible that the sharp peak around 79 $\mu$m in SNR N\,49 and the
PN or high-mass star IRAS\,05047$-$6644 is due to a blend of the cluster of
relatively strong transitions of H$_2$O and OH emission lines (Lerate et al.\
2006). Indeed, water vapour is produced in abundance in C-type shocks (Bergin,
Neufeld \& Melnick 1998), and is observed in YSO molecular outflows and SNRs
in the Galaxy (Bjerkeli et al.\ 2009). If true, this would suggest that both
water-dissociating and water-producing regions exist within the N\,49--ISM
interface. We find no evidence for molecular line emission in other MIPS-SED
spectra.

\subsubsection{Minerals}

Of the spectra classified with the suffix ``b'', the majority (13 out of 20)
do not show the [O\,{\sc iii}] line and the vast majority (16) have cold dust.
However, this is entirely consistent with the distribution of objects over
principal spectral types: there are 13 ``b'' sources among 35 objects without
[O\,{\sc iii}] --- i.e.\ $37(\pm10)$\% --- and there are 5 ``b'' sources among
13 objects with [O\,{\sc iii}] --- i.e.\ $38(\pm17)$\%; of 4 objects with
[O\,{\sc iii}] but not  [O\,{\sc i}], 2 are also a ``b'' source --- i.e.\
$50(\pm35)$\%. Likewise, 15 out of 30 ``C'' types bear suffix ``b'' --- i.e.\
$50(\pm13)$\% --- but so do 2 out of 8 ``W'' types --- i.e.\ $25(\pm18)$\%.
There is thus not a clear link between the spectral features responsible for
the unidentified bumps and the temperature and irradiation field. Rather than
a gaseous or cryogenic origin, a mineralogical explanation seems more likely.

A weak emission feature is seen around $\lambda\approx70$ $\mu$m, e.g., in the
warm (but neutral) envelopes of the LBV R\,71 and RSG IRAS\,05280$-$6910, and
perhaps in IRAS\,05137$-$6914. This might be due to crystalline forsterite
(Mg$_2$SiO$_4$), which is seen both in post- and pre-main sequence objects, at
$\lambda\approx69$--69.7 $\mu$m (Bowey et al.\ 2002); these would remain
unresolved in the MIPS-SED spectra. The similar mineral, crystalline
ortho-enstatite (MgSiO$_3$) has two peaks at either side of 70 $\mu$m
(Molster, Waters \& Tielens 2002), which together would appear as a single
feature in the MIPS-SED spectrum. Alternatively, chlorite (an
aluminum-containing hydrous silicate, which has nothing to do with chlorine)
peaks at $\lambda\approx69$ $\mu$m (Koike \& Shibai 1990); it also has a
strong peak at $\lambda\approx86$ $\mu$m, but that is so broad that it will be
difficult to recognise in the MIPS-SED spectrum.

Some objects display a bump around $\lambda\approx75$ $\mu$m, e.g., the YSO
N\,159S. An unidentified blend of weak, narrow features was also observed
around 74--75 $\mu$m in the PN NGC\,7027 (Liu et al.\ 1996), but the broad
feature we observe is more alike that seen in the high-mass star-forming core
G\,05.89 (Hatchell \& van der Tak 2003, who do not discuss it) --- the latter
also has an ice band around 66 $\mu$m, and no oxygen emission lines.

Bumps around $\lambda\approx78$ $\mu$m are seen, prominently in N\,89,
N\,51-YSO1, MSX-LMC\,577, and UFO\,1. These are all (candidate) YSOs. In the
Milky Way, HH\,53 shows a somewhat narrower feature (see Nisini et al.\
(1996), although it is not discussed). It sits on a cold continuum and only
the [O\,{\sc i}] is seen --- i.e.\ no (other) lines of water or OH, for
instance. It is thus unlikely to be due to o-H2O 4$_{23}$--3$_{12}$.

WOH\,G064, the source near WOH\,G457, and RP\,85 seem to have a bump nearer to
80 $\mu$m. This is also seen in the M0-type T\,Tauri star AA\,Tau and Herbig
Ae star MWC\,480 (Creech-Eakman et al.\ 2002).

These features remain unidentified, and it is not clear to what extent the
carriers of the features around 75, 78 and 80 $\mu$m are related.
Interestingly, serpentine (a magnesium/iron-containing hydrous silicate) peaks
at $\lambda\approx77$ $\mu$m (Koike \& Shibai 1990) and could be a viable
carrier for some of these observed spectral features.

Calcite (a carbonate) peaks at 90 $\mu$m and starts at 80 $\mu$m; it is seen
in the PNe NGC\,6302 (Kemper et al.\ 2002) and NGC\,6537 (Chiavassa et al.
2005) and possibly protostar NGC\,1333-IRAS\,4 (Ceccarelli et al.\ 2002).
Given that the feature is at the edge of the MIPS-SED spectral range it will
be difficult to convincingly argue for its presence in, for instance,
IRAS\,05328$-$6827 or BSDL\,2959.

\subsection{Trends in the line strengths and continuum slope}

\begin{figure*}
\epsscale{1.17}
\plotone{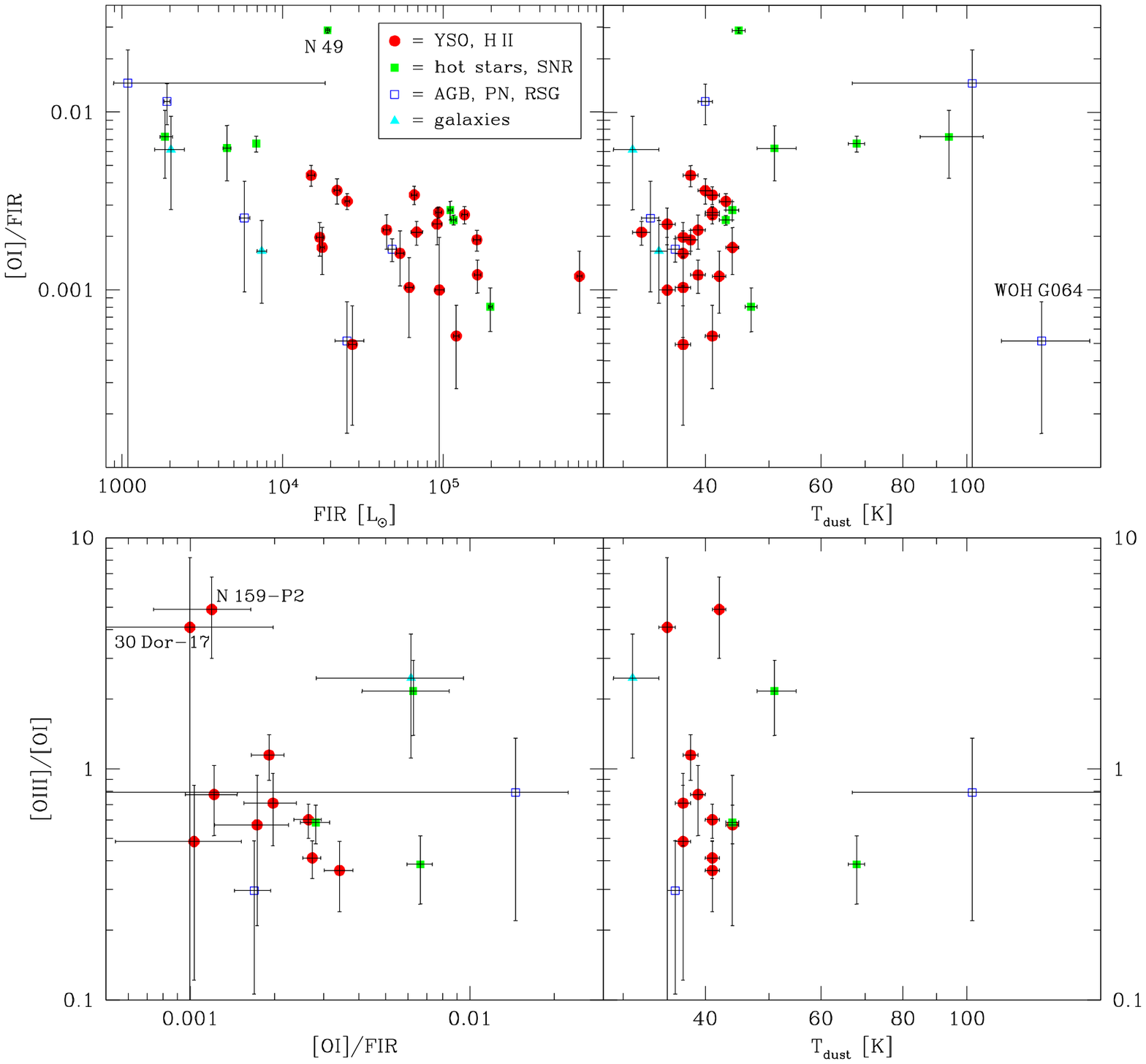}
\caption{Diagnostic diagrams utilizing the strength of the [O\,{\sc i}] and
[O\,{\sc iii}] fine-structure emission lines at $\lambda=63$ and 88 $\mu$m,
respectively, and the far-IR dust continuum (see text).}
\label{f7}
\end{figure*}

The MIPS-SED spectra can be described in terms of the oxygen line ratios and
line-to-continuum ratios, and the IR flux density and spectral slope. We use
these metrics to construct diagnostic diagrams, and investigate to what extent
these can be used to isolate sources of a particular nature (Fig.\ 7). To make
these diagrams as physically meaningful as possible, we make use of the dust
temperature derived from the spectral slope (\S 5.3.1); while for the continuum
we estimate the luminosity of the grey-body corresponding to that dust
temperature:
\begin{equation}
L({\rm FIR}) = 4 \pi d^2 \left.\frac{F_\lambda({\rm
observed})}{F_\lambda(T_{\rm dust})}\right|_{\rm ref}\ \int F_\lambda(T_{\rm
dust})\ {\rm d}\lambda,
\end{equation}
where we assumed a distance $d=50$ kpc and calibrated the grey-body (Eq.\ 3)
to the observed continuum interpolated across the oxygen lines at reference
wavelengths of 63 and 88 $\mu$m. Note that for a black-body, $\beta=0$ and the
integral yields the familiar $\sigma T_{\rm dust}^4$.

The far-IR luminosities derived in this way broadly agree with expectations;
for instance, it is always less than the stellar luminosity in the evolved
stars, but reaching significant fractions in the dusty RSG WOH\,G064
($\sim10$\%) and especially in the heavily-embedded RSG IRAS\,05280$-$6910
($>50$\%). The brightest YSOs and compact H\,{\sc ii} regions have far-IR
luminosities exceeding $10^5$ L$_\odot$, and these are likely powered by more
than one source, or additional luminosity is generated via accretion.

The clearest, if unsurprising, result arising from these diagrams (Fig.\ 7) is
that the objects associated with star formation (YSOs and compact H\,{\sc ii}
regions) separate quite well from the evolved stars by the former having
colder dust, $<44$ K. The few evolved stars that intrude into this cold regime
are likely to have swept-up ISM; evolved stars with unperturbed circumstellar
envelopes have dust of $\sim100$ K. No YSOs or compact H\,{\sc ii} regions are
seen to exhibit warm dust, but the embedded massive hot stars --- which are
likely unevolved --- occupy the regime of intermediate temperatures (44--100
K).

The [O\,{\sc iii}]/[O\,{\sc i}] ratio shows scatter over an order of
magnitude, but no relation to the line-to-IR ratio or dust temperature (Fig.\
7, bottom panels). This is in stark contrast to the clear correlation between
the [O\,{\sc iii}]/[O\,{\sc i}] ratio and dust temperature seen in H\,{\sc ii}
regions in M\,33 (Higdon et al.\ 2003). Two objects stand out, viz.\ N\,159-P2
and 30\,Dor-17, with their very strong [O\,{\sc iii}] lines.

In PDRs, the [O\,{\sc i}] line is the dominant cooling line for high density
gas. The IR luminosity is a measure of the input energy into the molecular
cloud surface (the dust acts as a calorimeter). The ratio of the [O\,{\sc i}]
line to IR luminosity is therefore an approximate measure of the efficiency of
the photo-electric heating process which is thought to dominate the heating in
these PDRs. This efficiency appears to be $\approx0.1$--0.3\% for most of the
YSOs and compact H\,{\sc ii} regions (Fig.\ 7, top left panel, in which N\,49
stands out for its shock-excited [O\,{\sc i}] line).

\section{Conclusions}

We have presented the 52--93 $\mu$m spectra of 48 compact far-IR sources in
the Large Magellanic Cloud, obtained with MIPS-SED onboard the {\it Spitzer
Space Telescope} as part of the SAGE-Spec Legacy Program. The spectra were
classified using a simple classification scheme introduced here for the first
time. We measured the intensity of the fine-structure lines of oxygen ---
[O\,{\sc i}] at 63 $\mu$m and [O\,{\sc iii}] at 88 $\mu$m --- as well as the
slope of the dust continuum emission spectrum which we translated into a dust
temperature and far-IR luminosity. Some of the most interesting results
arising from this analysis may be summarized as follows:

\begin{itemize}
\item[$\bullet$]{Young Stellar Objects (YSOs) separate rather well from more
evolved objects by the steeper slope of the far-IR spectrum (i.e.\ lower dust
temperature) and to some extent brighter fine-structure line emission. The
contributions from [O\,{\sc i}] and [O\,{\sc iii}] show large variations among
the sample, likely due to different excitation mechanisms at play.}
\item[$\bullet$]{The supernova remnant N\,49 displays spectacular [O\,{\sc i}]
line emission, contributing 11\% to the MIPS 70-$\mu$m broad-band flux. We
interpret this as arising from shocked gas at the interface of the expanding
supernova ejecta and the local ISM. A possible detection of water vapour
and/or OH emission is reported. From the dust continuum emission we estimate
that 0.2 M$_\odot$ of dust is associated, consistent with the expected amount
of collected ISM.}
\item[$\bullet$]{Equally spectacular emission is seen in the [O\,{\sc iii}]
line in 30\,Dor-17 and N\,159-P2. These objects are associated with regions of
active star formation and are probably YSOs harboring an (ultra-)compact
H\,{\sc ii} region.}
\item[$\bullet$]{The ratio of [O\,{\sc i}] to IR (dust-processed) luminosity
is used to estimate the efficiency of photo-electric heating in the interfaces
between ionized gas and molecular clouds in YSOs and compact H\,{\sc ii}
regions; we find it is $\approx0.1$--0.3\%.}
\item[$\bullet$]{We derive a low nitrogen content of H\,{\sc ii} regions in
the LMC, with a nitrogen-to-oxygen ratio of $N({\rm N})/N({\rm O})<0.1$,
possibly $<0.02$. This confirms the sparse evidence previously available from
far-IR fine-structure lines.}
\item[$\bullet$]{The [O\,{\sc i}] line is detected in the extreme red
supergiant WOH\,G064. It is strong compared to the line detected in Galactic
red supergiants, hinting at a larger dust-free cavity at lower metallicity.
The implication may be a larger contribution of Alfv\'en waves to the wind
driving at lower metallicity.}
\item[$\bullet$]{The circumstellar envelopes of the OH/IR stars in our sample
lack cold dust. We estimate that dust production must have increased
dramatically only a few thousands years ago, for the red supergiants and
massive AGB stars alike. We further estimate that the {\it fractional} mass
lost in this ``superwind'' phase is rather similar for these stars in spite of
their different birth masses.}
\item[$\bullet$]{The Luminous Blue Variable R\,71 also lacks cold dust, but
the timescales are somewhat longer and the coolest dust may have formed a few
$10^4$ yr ago; this would be consistent with an origin in the wind of a red
supergiant progenitor to R\,71, as suggested previously by others.}
\item[$\bullet$]{In the case of the carbon stars in our sample, which show
far-IR excess emission over that expected from a circumstellar shell, and
possibly also some planetary nebulae, we interpret the far-IR emission as
arising from swept-up ISM and not due to dust produced by the stars
themselves.}
\item[$\bullet$]{Broad (several $\mu$m wide) emission features are seen in
many sources, both young and evolved. It is likely that these arise from modes
within solid state material, probably minerals, but the exact carriers could
not be identified (maybe hydrous silicates).}
\item[$\bullet$]{Emission in the 60--70 $\mu$m region due to crystalline water
ice is detected tentatively in several YSOs.}
\end{itemize}

\acknowledgments

We thank Chad Engelbracht for commenting on Paper II, which also helped
improve this Paper I. We also thank the anonymous referee for her/his
favorable report.


\clearpage
\LongTables
\begin{landscape}

\begin{deluxetable*}{p{3mm}lllll}
\tabletypesize{\scriptsize}
\tablecaption{Description of compact sources in the LMC as targets for
MIPS-SED.}
\tablehead{
\colhead{\#}                                  &
\colhead{Principal name\ \tablenotemark{a}}   &
\colhead{Alternative name\ \tablenotemark{a}} &
\colhead{Object type\ \tablenotemark{b}}      &
\colhead{RA and Dec (J2000)}                  &
\colhead{References}                          }
\startdata
1                        &
SAGE\,04374$-$6754\rlap{\ \tablenotemark{c}} &
                         &
(Galaxy)                 &
4 37 27.61 $-$67 54 35.0 &
none                     \\
2                        &
SMP-LMC\,11              &
IRAS\,04515$-$6710       &
proto-PN                 &
4 51 37.82 $-$67 05 17.0 &
9,10,16,38,39,42,71,94,95,100,122,123,147,153,205,222,229 \\
3                        &
IRAS\,04530$-$6916       &
                         &
B[e], YSO ?              &
4 52 45.70 $-$69 11 49.4 &
30,42,47,51,63,84,100,155,170,193,195,196,206,214,223,230 \\
4                        &
IRAS\,04537$-$6922       &
Brey\,3a                 &
WR ?                     &
4 53 30.11 $-$69 17 49.3 &
21,42,63,66,67,84,98,100,115,119 \\
5                        &
N\,89                    &
LM\,1-8                  &
VLE                      &
4 55 06.53 $-$69 17 08.5 &
42,63,79,84,98,100,146   \\
6                        &
WOH\,G064                &
IRAS\,04553$-$6825       &
RSG, OH/IR               &
4 55 10.48 $-$68 20 29.9 &
23,42,43,84,90,97,100,104,128,144,170,174,188,190,192,193,195,196,199,213 \\
 &
 &
 &
 &
 &
214,221,223,230 \\
7                        &
IRAS\,04557$-$6639       &
[HKN2006]\,25            &
HAeBe                    &
4 55 50.59 $-$66 34 34.7 &
42,61                    \\
\enddata
\end{deluxetable*}

\clearpage
\end{landscape}

\begin{deluxetable}{p{3mm}lcccl}
\tablecaption{Description of MIPS-SED data of compact sources in the LMC.}
\tablehead{
\colhead{\#}                               &
\colhead{AOR Key}                          &
\colhead{Integration\ \tablenotemark{a}}   &
\colhead{Quality}                          &
\colhead{Extraction}                       &
\colhead{Notes}                            }
\startdata
1                  &
22459648           &
\llap{20}\ $\times$\ \rlap{10} &
good               &
on--off            &
                   \\
2                  &
22450432           &
\llap{20}\ $\times$\ \rlap{10} &
ok                 &
on--off            &
                   \\
3                  &
22453504           &
\llap{10}\ $\times$\ \rlap{\hspace{0.5mm} 3} &
good               &
on--off            &
                   \\
4                  &
22457088           &
\llap{20}\ $\times$\ \rlap{10} &
good               &
on--off            &
                   \\
5                  &
22460416           &
\llap{8}\ $\times$\ \rlap{\hspace{0.5mm} 3} &
good               &
on--off            &
                   \\
6                  &
22448640           &
\llap{8}\ $\times$\ \rlap{10} &
good               &
on--off            &
                   \\
7                  &
22457344           &
\llap{3}\ $\times$\ \rlap{10} &
good               &
on--off            &
                   \\
8                  &
22457600           &
\llap{6}\ $\times$\ \rlap{10} &
good               &
on--off            &
                   \\
\enddata
\tablenotetext{a}{Total on-source integration time, $N_{\rm cycles}\times
t$(s).}
\tablenotetext{b}{Not detected.}
\tablenotetext{c}{Extended 63-$\mu$m line emission.}
\tablenotetext{d}{Spectrum of a bright point source in offset position.}
\tablenotetext{e}{Offset position has high background.}
\tablenotetext{f}{Spectrum of a bright point source in offset position.}
\tablenotetext{g}{Spectrum of a second bright point source on the slit.}
\tablenotetext{h}{Extended continuum emission? See also Figure 3.}
\tablenotetext{i}{Offset position has high background.}
\tablenotetext{j}{Extended continuum and 88-$\mu$m line emission.}
\tablenotetext{k}{Very bright point source in the offset position at the edge
of the slit.}
\tablenotetext{l}{Spectrum of a bright point source in offset position.}
\tablenotetext{m}{Spectrum of a bright point source in offset position
$\equiv$ MSX-LMC\,956.}
\tablenotetext{n}{Spectrum of a bright point source in offset position
$\equiv$ SAGE\,05407$-$7011.}
\tablenotetext{o}{Spectrum of a bright point source in offset position,
centered outside the slit (Fig.\ 3).}
\end{deluxetable}

\clearpage

\begin{deluxetable*}{p{3mm}lrrrrrrrl}
\tablecaption{MIPS photometric data of MIPS-SED targets in the LMC.}
\tablehead{
\colhead{\#}                       &
\colhead{AOR Target}               &
\colhead{$F_\nu(24)$}              &
\colhead{$\sigma(24)$}             &
\colhead{$F_\nu(70)$}              &
\colhead{$\sigma(70)$}             &
\colhead{$F_\nu(160)$}             &
\colhead{$\sigma(160)$}            &
\colhead{$Var$\ \tablenotemark{a}} &
\colhead{Notes}                    \\
                                   &
                                   &
[mJy]                              &
[mJy]                              &
[mJy]                              &
[mJy]                              &
[mJy]                              &
[mJy]                              &
                                   &
                                   }
\startdata
1                  &
SAGE\,04374$-$6754 &
11.1               &
0.2                &
200                &
4                  &
                   &
                   &
0                  &
                   \\
2                  &
SMP-LMC\,11        &
489.5              &
1.8                &
108                &
3                  &
                   &
                   &
4                  &
                   \\
3                  &
IRAS\,04530$-$6916 &
3251.5             &
11.9               &
15110              &
287                &
15420              &
290                &
38                 &
                   \\
4                  &
IRAS0\,4537$-$6922 &
1295.0             &
4.5                &
768                &
19                 &
                   &
                   &
0                  &
                   \\
5                  &
N\,89              &
1013.9             &
3.3                &
8725               &
110                &
5908               &
237                &
11                 &
                   \\
6                  &
WOH\,G064          &
865.2              &
17.2               &
1954               &
38                 &
                   &
                   &
                   &
b                  \\
\enddata
\tablenotetext{a}{The variability at 24 $\mu$m is defined as:
$Var \equiv 100\times\left(F_{\nu,\ {\rm max}}/F_{\nu,\ {\rm min}}-1\right)$.}
\tablenotetext{b}{Only one 24-$\mu$m epoch.}
\tablenotetext{c}{No 24-$\mu$m data.}
\tablenotetext{d}{There is another source at $9^{\prime\prime}$ with
$F_\nu(24)=123/113$ mJy.}
\tablenotetext{e}{The associated 70-$\mu$m source is at $10^{\prime\prime}$.}
\tablenotetext{f}{The associated 24-$\mu$m source is at $8^{\prime\prime}$,
and there is another source at $10^{\prime\prime}$ with $F_\nu(24)=207$ mJy.}
\tablenotetext{g}{The associated 70-$\mu$m source is at $15^{\prime\prime}$,
the associated 160-$\mu$m source is at $25^{\prime\prime}$, and there is
another source at $6^{\prime\prime}$ with $F_\nu(24)=69$ mJy.}
\tablenotetext{h}{The associated 70-$\mu$m source is at $6^{\prime\prime}$,
and there is another source at $13^{\prime\prime}$ with $F_\nu(70)=50$ mJy.}
\tablenotetext{i}{The associated 70-$\mu$m source is at $14^{\prime\prime}$,
but the associated 160-$\mu$m source is at $10^{\prime\prime}$.}
\tablenotetext{j}{The associated 70-$\mu$m source is at $5^{\prime\prime}$,
and there is another source at $18^{\prime\prime}$ with $F_\nu(70)=88890$ mJy.
At epoch 2, there was another source at $9^{\prime\prime}$ with
$F_\nu(24)=1253$ mJy.}
\tablenotetext{k}{At epoch 1, there was another source at $4^{\prime\prime}$
with $F_\nu(24)=33$ mJy.}
\tablenotetext{l}{The associated 70-$\mu$m source is at $19^{\prime\prime}$.}
\tablenotetext{m}{There is another source at $16^{\prime\prime}$ with
$F_\nu(24)=46$ mJy, and at epoch 2 there was yet another source at
$7^{\prime\prime}$ also with $F_\nu(24)=46$ mJy. For comparison: the 70-$\mu$m
source associated with the MIPS-SED target is at only $4^{\prime\prime}$.}
\end{deluxetable*}

\clearpage

\begin{deluxetable*}{p{3mm}llclccrl}
\tablecaption{Quantities derived from the MIPS-SED spectra of compact sources
in the LMC.}
\tablehead{
\colhead{\#}                                &
\colhead{AOR Target}                        &
\colhead{Type\ \tablenotemark{a}}           &
\colhead{$\alpha$\ \tablenotemark{b}}       &
\colhead{$T_{\rm dust}$\ \tablenotemark{c}} &
\colhead{$L({\rm [O\,I]})$}                 &
\colhead{$L({\rm [O\,III]})$}               &
\colhead{$L({\rm FIR})$\ \tablenotemark{d}} &
\colhead{Notes}                             \\
                                            &
                                            &
                                            &
                                            &
[K]                                         &
\multicolumn{3}{c}{---------------------\ \ \ [L$_\odot$]\ \ \ ---------------------} &
                                            }
\startdata
1                  &
SAGE\,04374$-$6754 &
C0                 &
$1.11\pm0.21$      &
$31_{-2}^{+3}$     &
\llap{1}$5\pm6$    &
\llap{3}$1\pm7$    &
$2.0_{+0.4}^{-0.4}\times10^3$ &
                   \\
2                  &
SMP-LMC\,11        &
C0                 &
$0.44\pm0.09$      &
$40_{-1}^{+1}$     &
\llap{2}$2\pm6$    &
$<9$               &
$1.91_{+0.09}^{-0.08}\times10^3$ &
                   \\
3                  &
IRAS\,04530$-$6916 &
C1b                &
$0.370\pm0.010$    &
$41_{-1}^{+1}$     &
$67\pm33$          &
$<4$\rlap{2}       &
$1.21_{+0.05}^{-0.05}\times10^5$ &
                   \\
4                  &
IRAS0\,4537$-$6922 &
W2                 &
\llap{$-$}$0.638\pm0.035$ &
$68_{-2}^{+2}$     &
\llap{4}$6\pm5$    &
\llap{1}$8\pm5$    &
$6.878_{+0.005}^{+0.011}\times10^3$ &
                   \\
5                  &
N\,89              &
C2b                &
$0.345\pm0.019$    &
$41_{-1}^{+1}$     &
\llap{2}$56\pm16$  &
\llap{1}$06\pm18$  &
$9.39_{+0.4}^{-0.35}\times10^4$ &
                   \\
6                  &
WOH\,G064          &
W1b                &
\llap{$-$}$1.20\pm0.08$  &
\llap{1}$30_{-17}^{+24}$ &
\llap{1}$3\pm8$    &
$<1$\rlap{2}       &
$2.53_{-0.39}^{+0.7}\times10^4$ &
                   \\
\enddata
\tablenotetext{a}{The spectral classification scheme is described in \S 4.3.}
\tablenotetext{b}{The spectral slope is defined as: $\alpha \equiv 2.44\
\frac{F_\nu(85)-F_\nu(55)}{F_\nu(85)+F_\nu(55)}$.}
\tablenotetext{c}{The dust temperature was estimated from the spectral slope
as described in \S 5.3.1 (cf.\ Fig.\ 6).}
\tablenotetext{d}{The far-IR luminosity was estimated based on the dust
temperature as described in \S 5.5, assuming a distance of 50 kpc.}
\tablenotetext{e}{Possible broad crystalline water ice band between
$\lambda=60$--70 $\mu$m.}
\tablenotetext{f}{Possible blend of molecular lines of water and/or hydroxyl
at $\lambda\approx79$ $\mu$m.}
\end{deluxetable*}

\clearpage

\begin{deluxetable*}{p{3mm}llclcclc}
\tablecaption{Description of serendipitous MIPS-SED data in the LMC.}
\tablehead{
\colhead{\#}                                &
\colhead{Identification\ \tablenotemark{a}} &
\colhead{Object type\ \tablenotemark{b}}    &
\colhead{RA and Dec (J2000)}                &
\colhead{References}                        &
\colhead{Quality}                           &
\colhead{Extraction}                        &
\colhead{Type\ \tablenotemark{c}}           &
\colhead{$\alpha$\ \tablenotemark{d}}       }
\startdata
16B                   &
SAGE\,05138$-$6935\rlap{\ \tablenotemark{e}} &
(H\,{\sc ii} ?)       &
5 13 46.5 $-$69 35 10 &
none                  &
poor                  &
off                   &
C0                    &
$1.04\pm0.05$         \\
23B                   &
N\,49                 &
SNR                   &
5 26 07   $-$66 05.0  &
(Table 1)             &
poor                  &
off                   &
C1                    &
$0.760\pm0.034$       \\
24B                   &
LH\,55                &
NC                    &
5 26 38.6 $-$67 39 23 &
3,8,11                &
ok                    &
off                   &
F0                    &
$0.74\pm0.20$         \\
30B                   &
IRAS\,05330$-$6826    &
H\,{\sc ii}           &
5 32 48.3 $-$68 23 59 &
2,7,9,10              &
good                  &
off                   &
C2                    &
$0.473\pm0.033$       \\
31B                   &
SAGE\,05328$-$6930\rlap{\ \tablenotemark{f}} &
(H\,{\sc ii} ?)       &
5 32 49.4 $-$69 30 37 &
none                  &
poor                  &
on                    &
C2                    &
$0.88\pm0.09$         \\
38B                   &
[H88b]\,86 ?          &
Mol ?                 &
5 38 40.9 $-$69 25 14 &
5                     &
poor                  &
off                   &
C3                    &
$0.83\pm0.05$         \\
\enddata
\tablenotetext{a}{Names of the type ``N\,[number]'' are
``LHA\,120-N\,[number]'' in full.}
\tablenotetext{b}{See the explanation of Table 1.}
\tablenotetext{c}{The spectral classification scheme is described in \S 4.3.}
\tablenotetext{d}{The spectral slope is defined as: $\alpha \equiv 2.44\
\frac{F_\nu(85)-F_\nu(55)}{F_\nu(85)+F_\nu(55)}$.}
\tablenotetext{e}{Abbreviation of the blend of MIPS 24-$\mu$m sources
SSTM1SAGE1\,J051347.73$-$693505.3 and SSTM1SAGE1\,J051344.98$-$693510.2
(following the IRAS convention).}
\tablenotetext{f}{Abbreviation of SSTM1SAGE1\,J053249.35$-$693036.8 (following
the IRAS convention).}
\tablenotetext{g}{At the edge of the slit.}
\tablenotetext{h}{Abbreviation of SSTISAGE1C\,J054043.15$-$701110.3 (following
the IRAS convention).}
\tablerefs{
1.  Bica et al.\ (1999);
2.  Egan et al.\ (2001);
3.  Filipovi\'c, Jones \& White (2003);
4.  Gatley, Hyland \& Jones (1982);
5.  Hodge (1988);
6.  Johansson et al.\ (1998);
7.  Kastner et al.\ (2008);
8.  Lucke \& Hodge (1970);
9.  Marx, Dickey \& Mebold (1997);
10. van Loon et al.\ (2005c);
11. Wang \& Helfand (1991).
}
\end{deluxetable*}

\end{document}